# Broadband Transient Full-Stokes Luminescence Spectroscopy with High Sensitivity

Antti-Pekka M. Reponen[1], Marcel Mattes[1,2], Zachary A. VanOrman[1,2], Lilian Estaque[3], Grégory Pieters[3], and Sascha Feldmann[1,2]*

[1]Rowland Institute, Harvard University, Cambridge, MA, USA
[2]Institute of Chemical Sciences and Engineering, École Polytechnique Fédérale de Lausanne, Lausanne, Switzerland
[3]Université Paris-Saclay, CEA, INRAE, Département Médicaments et Technologies pour la Santé; Gif-sur-Yvette, France

*Email: sascha.feldmann@epfl.ch

## Abstract

Materials emitting circularly polarized light (CPL) are highly sought after for applications ranging from efficient displays to quantum information technologies. Established methods for time-resolved CPL characterization have significant limitations, preventing in-depth photophysical insight necessary for materials development. We have designed and built a high-sensitivity (noise level $10^{-4}$), broadband (ca. 400-900 nm), transient (ns resolution, ms range) full-Stokes (CPL and linear polarizations) spectroscopy setup. We demonstrate its broad applicability by measuring compounds with low dissymmetry factors across various timescales, as well as tracking the temporal evolution of linear polarization components alongside associated CPL artifacts. We have written open-source software and used stock optical components to make transient CPL spectroscopy practically accessible to a wide audience, enabling the study of chiral materials in numerous diverse applications.

# Main

Circularly polarized luminescence (CPL) emitting materials have seen a resurgence in research interest over recent years. This is largely driven by emerging and commercially promising technologies utilizing CPL in photonics and optoelectronics, spanning a broad range from photonic and quantum computing[1,2], security inks[3], efficient device displays[4] as well as holographic[5] and 3D display[6] technologies. Accordingly, there is currently significant interest in accurate characterization of circularly polarized light emission processes.

Earliest CPL measurements trace back to the mid-20$^{th}$ century[7,8], and in the following decades high-sensitivity CPL instruments were developed; using a photoelastic modulator (PEM) for rapid polarization modulation[9,10] followed by lock-in amplification (LIA) or differential photon counting (DPC) relative intensity differences of order $10^{-5}$ are measurable. These basic detection principles continue to underlie most high-sensitivity CPL measurements today. CPL is often characterized by the dissymmetry factor, defined as

$$g_{lum} = \frac{\Delta I_{LCP/RCP}}{I_{avg}} = \frac{I_{LCP} - I_{RCP}}{\frac{1}{2}(I_{LCP} + I_{RCP})}$$

where $I_{LCP}$ and $I_{RCP}$ refer to intensity of LCP (left-handed circularly polarized) and RCP (right-handed circularly polarized) light, respectively.

Although time-resolved luminescence measurements are ubiquitous in the characterization of emissive materials, translating CPL measurements to the time domain remains rare. This is despite the multiple time-resolved CPL (TRCPL) instruments based on photoelastic modulation already described in the 1990s[11-15] alongside their applications in revealing racemization and energy transfer dynamics[16,17]. In particular, Schauerte et al. showed in 1995 that combining DPC with time-correlated single photon counting (TCSPC) can achieve ns time resolution and accurate measurement of dissymmetry values at the $10^{-3}$ level[18].

In practice, combining photoelastic modulation with time resolution introduces severe limitations. The modulation rate must be compatible with the detector readout rate, the excitation repetition rate, and the timescale of the luminescence decay. As the detection scheme is not broadband, spectra must be built up one wavelength point at a time. Even in the steady state, full spectrum acquisition is time-consuming[19]. With TCSPC additionally dividing the luminescence signal into multiple time bins and limiting detector count rates to ~1-5% of the excitation rate to avoid photon pile-up, collecting time-resolved CPL spectra becomes impractical (more discussion in SI Section 2.6). We note that Schauerte et al. reported long measurement times despite using MHz excitation rates, which already excludes long-lived decay processes. In addition, using a simple filter to select the emission range[18] prevented the collection of spectral information.

Such limitations may be why the few contemporary works in TRCPL[20-23] rely on quarter-wave plates (QWP) instead of modulation, resulting in simpler instruments and faster acquisition. However, the lack of modulation results in low CPL sensitivity, and the timescales covered are in the μs-ms range. These approaches are therefore practically limited to chiral lanthanide complexes with extreme dissymmetries and long-lived emission[24].

However, lanthanide complexes only make up a fraction of the diverse material platforms that are currently explored for CPL applications, many of which feature short-lived (nanosecond) luminescence and/or small dissymmetry factors (~$10^{-3}$)[25,26]. Hence, to fuel these exciting developments, TRCPL methods are required that combine sensitive broadband acquisition with the flexibility to probe both fluorescence (ns) and phosphorescence (µs-ms) on their respective timescales.

*Measuring principle*

Our novel approach to TRCPL is summarized in Figure 1. It makes use of an electronically gated CCD camera, which allows for flexible time gates from 2 ns to several ms, is inherently broadband, and is not limited by photon pile-up effects. To achieve high sensitivity without polarization modulation, we simultaneously record orthogonal linear polarizations on different parts of the CCD array (2048×512 pixels), as recently applied by Baguenard *et al.* for calibration-free error-cancelling in a steady-state CPL spectrometer[27] and previously used in fluorescence correlation time imaging[28]. The reader is referred to those works for an excellent coverage of the theory and associated errors. Error cancelling enables sensitive measurements of dissymmetry, with noise floors on the order of $10^{-4}$ for sufficiently bright emission. Using a single CCD additionally eliminates issues with channel synchronicity and reduces overall cost and complexity compared to dual-detector approaches[23].

Spatial separation of orthogonal polarizations is achieved by a Wollaston prism, resulting in a free-space setup with no fiber coupling or mirrors between the sample and spectrograph. To separate circularly polarized luminescence components rather than linear ones, a QWP with its fast axis at 45⁰ to the Wollaston axes is placed in front of the prism.

Cancellation of errors due to different transmissions along the two beam paths and time instability is achieved by repeating the measurement with the QWP rotated by 90⁰ (automated by a precision motorized mount), which swaps the two channels on the CCD detector[27]. Analogously, for quantifying linear polarization (LP) components (a primary error source in CPL measurements[19]), we additionally incorporate a half-wave plate (HWP), rotation of which allows swapping the emitted LP components passed to each channel. This results in a sensitive, time-resolved broadband measurement of the full Stokes vector. Waveplate interactions with various polarization components are shown in Fig. 1c. At suitable angle combinations, only a specific component of the Stokes vector is translated to the intensity difference between channels.

Sample excitation is possible in various configurations, defined by the relative angle and polarization of the excitation beam with respect to the collection optics. The subtleties of the excitation geometry have been well described by Blok and Dekkers[29]. Photoselection effects can be especially relevant in time-resolved studies, if relaxation of dipole orientations occurs slower than the measurement time resolution[30].

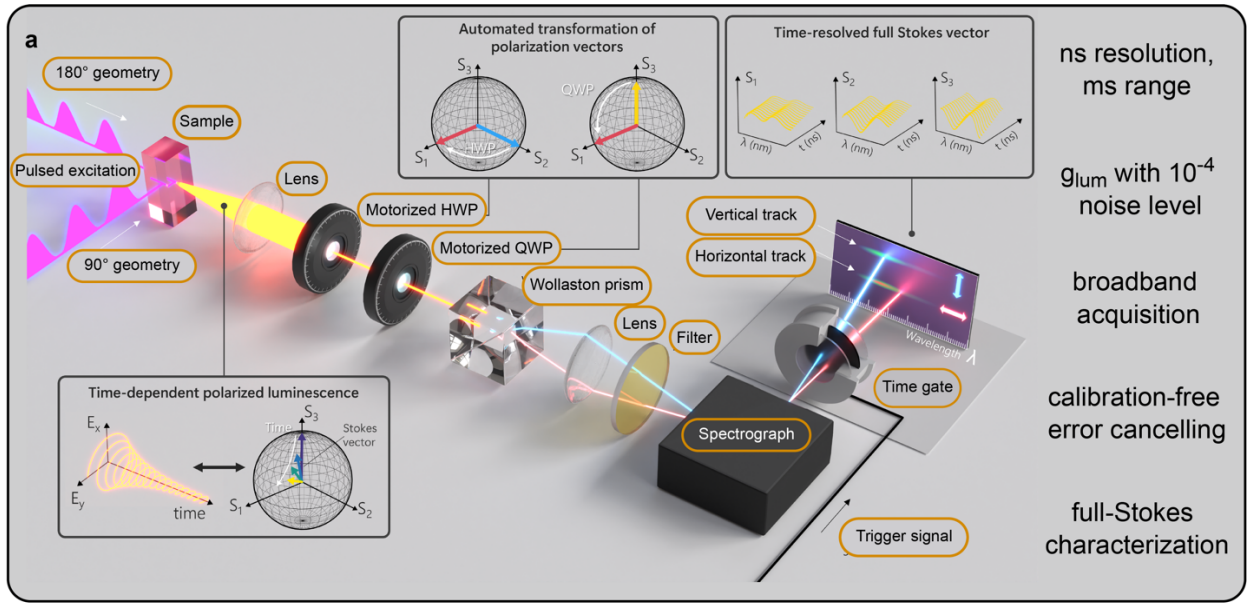
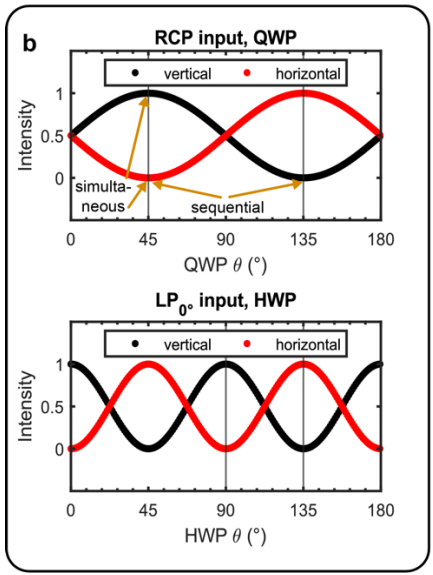
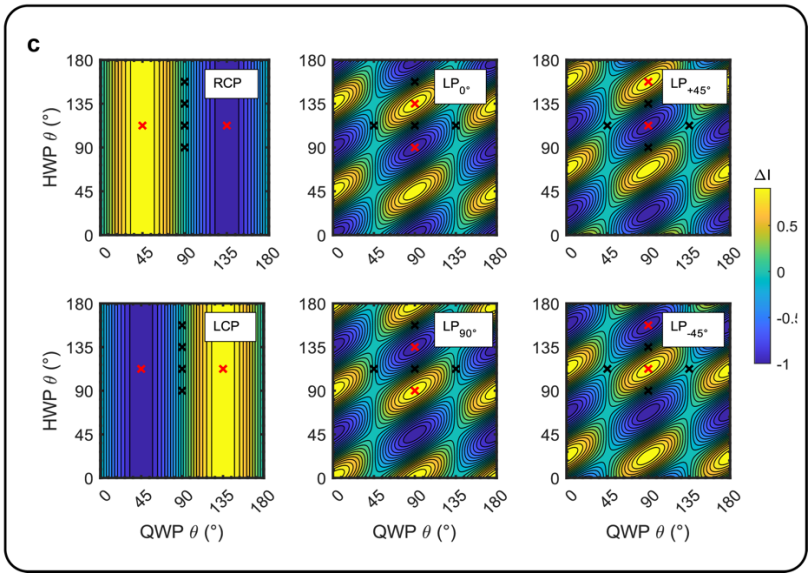

**Fig. 1 | Overview of key setup components, features and principles. a**, Schematic diagram of the key optical components and measurement procedure. Excitation by pulsed laser (200 fs, variable repetition rate 0.5-50 kHz) in the 90º geometry eliminates certain polarization artifacts, while the 180º geometry is more suitable for thin samples. Super-achromatic waveplates (HWP for half-wave plate, QWP for quarter-wave plate) in appropriate orientations transform polarization components into orthogonal linear components, which are separated by a Wollaston prism and passed through a grating spectrograph. Orthogonally polarized spectra are recorded simultaneously as two tracks on a single CCD array detector. An intensifier tube is electronically gated to provide time resolution. Waveplate rotation, automatable by motorized housings, allows swapping the detection tracks which results in cancelling of the largest channel transmission mismatches and temporal instabilities. **b**, Calculated intensities at vertical and horizontal detection

tracks as a function of QWP angle for a right-handed circularly polarized (RCP) luminescence input and HWP angle for a linearly polarized input. Two tracks are simultaneously recorded, and the measurement repeated at a second QWP/HWP position where tracks swap. **c**, Calculated intensity difference between measurement tracks as a function of both QWP and HWP angle for the pure Stokes basis polarizations. Red crosses indicate angles where measurements are taken for that polarization component. For reference, black crosses (each of which is at ΔI = 0) indicate angles where measurements are taken for other polarization components.

All data presented in this work was collected from solution samples, for which an orthogonal geometry with square-based four-window cuvettes is the most straightforward approach. In this geometry, photoselection effects can be minimized by using a horizontal excitation polarization and maximized by a vertical excitation polarization. Besides photoluminescence, circularly polarized electroluminescence (CP-EL), *e.g.*, from a spin-LED, could form the light source to be characterized with suitable modifications to excitation. For more details on the setup, methodology and analysis we refer the reader to the Supporting Information (Sections 1 & 2).

For the remainder of this work, we showcase setup performance with data collected on small molecules in solution with varying degrees (strong, weak, none) of CPL dissymmetry and excited-state timescales from single-ns to >100 μs.

### Chiral lanthanide complex

*Steady state and microsecond gating of Eu[(+)-facam]$_3$ luminescence*

Lanthanide complexes with chiral ligands often show strong CPL[24,31]. In particular, Eu[(+)-facam]$_3$ ((+)-facam = 3-(trifluoromethylhydroxymethylene)-(+)-camphorate; structure shown in Fig. 2) is commercially available and possesses high emission dissymmetry ($|g_{lum}|$ up to 0.78 in DMSO at 595 nm)[9]. As such, Eu[(+)-facam]$_3$ is a common standard for CPL setup validation and testing for which multiple literature sources exist for comparison[9,19]. Such reports include a recent CCD-based CPL setup with time gating functionality[23] and fully time-resolved studies by Hananel[22], which shows the temporal evolution of dissymmetry with approximately 10 μs resolution and a millisecond range.

To establish comparison with existing literature, we first present steady-state and long timescale time-resolved (50 μs gate width) CPL spectra of Eu[(+)-facam]$_3$ in dry DMSO (Fig. 2). The sharp emission features in this complex arise from f-f transitions on the $Eu^{3+}$ centre, which generally are of the form $^5D_{J1} \rightarrow {}^7F_{J2}$ (where $J_1$ and $J_2$ are the total angular momentum quantum numbers) and can be assigned to $^5D_0 \rightarrow {}^7F_{J2}$ transitions specifically[32-34]. Of these, the most striking CPL features appear at 595 nm and 613 nm, corresponding to a $^5D_0 \rightarrow {}^7F_1$ magnetic dipole transition and a $^5D_0 \rightarrow {}^7F_2$ induced electric dipole transition, with reported $g_{lum}$ values of –0.78 and +0.072, respectively[9,19,22] (albeit with some variance and environmental sensitivity[35], notably to water[22,36]).

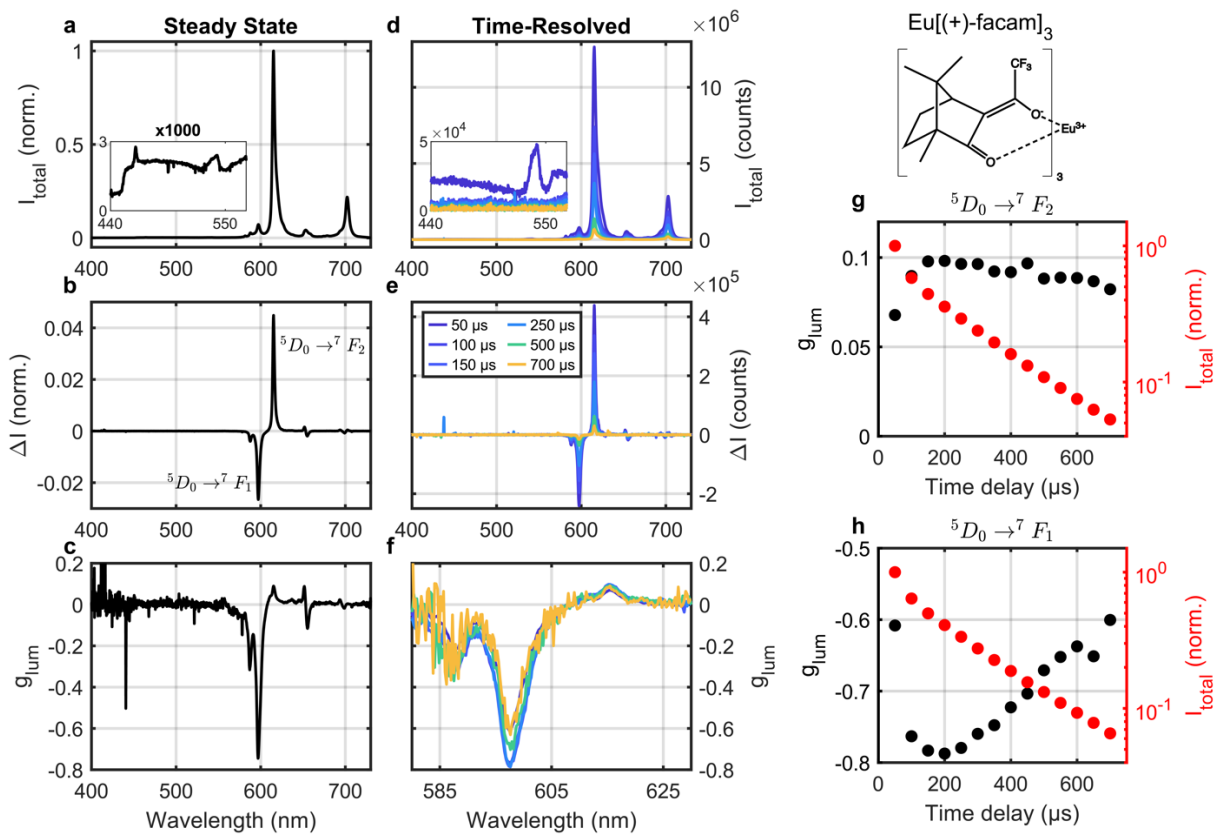

**Fig. 2 | Steady-state and microsecond time-resolved CPL spectroscopy of the chiral standard Eu[(+)-facam]$_3$ (structure shown top right) in DMSO solution. a,** Total steady-state luminescence intensity (405 nm CW excitation, 450 nm long-pass filter). The inset shows weak emission on the blue side, partially cut off by the filter. **b,** Steady-state CPL spectrum showing the intensity difference $\Delta I$ between LCP and RCP luminescence, normalized to the maximum total luminescence intensity. The main CPL features are labelled, $^5D_0\rightarrow{}^7F_1$ near 595 nm and $^5D_0\rightarrow{}^7F_2$ near 615 nm. **c,** Steady-state spectrum of the luminescence dissymmetry factor $g_{lum}$. **d,** Time-resolved total luminescence spectra (343 nm excitation pulsed at 1 kHz, 380 nm long-pass, 50 μs time bins) at selected gate delays. **e,** Time-resolved CPL spectra at selected gate delays. **f,** Time-resolved spectra of the luminescence dissymmetry factor at selected gate delays. **g,** Intensity and dissymmetry factors as a function of time at the $^5D_0\rightarrow{}^7F_2$ peak. **h,** Intensity and dissymmetry factors as a function of time at the $^5D_0\rightarrow{}^7F_1$ peak (small upticks at 450 μs/650 μs in **g/h**, respectively, are likely due to cosmic rays).

In the steady state, our luminescence and CPL spectra match literature (Fig. 2a-c), although the recorded $g_{lum}$ of -0.745 for the $^5D_0\rightarrow{}^7F_1$ transition is slightly lower than expected. Besides potential issues with, *e.g.*, water ingress, a possible explanation is the relatively coarse grating necessary for broadband acquisition and finite detector pixel size resulting in a wavelength resolution limit of

~1 nm, similar to the line width of the feature (see SI Fig. S10 for the effects of slit width on measured $g_{lum}$).

While the time-resolved luminescence spectra (Fig. 2d-f) show little evolution of the spectral shape over time, the magnitude of $g_{lum}$ changes over time as the emission decays (Fig. 2g-h). The reduction in $g_{lum}$ over time is similar to what was observed in the Hananel *et al.* study[22], where a decrease from a peak value of approximately -0.8 with a time constant of approximately 1 ms is reported and assigned to sample heterogeneity. Intriguingly, our data appears to show an increase in $g_{lum}$ for the first few time bins. This is more apparent for the $^5D_0 \rightarrow {}^7F_1$ peak (Fig. 2h) than the $^5D_0 \rightarrow {}^7F_2$ peak (Fig. 2g). The early-time rise in $g_{lum}$ is likely linked to weak transitions seen on the blue side in the earliest time bins (Fig. 2d, inset) which will overlap more with the $^5D_0 \rightarrow {}^7F_1$ peak at 595 nm than the $^5D_0 \rightarrow {}^7F_2$ peak at 613 nm. Such features are also faintly visible in the steady-state measurement (Fig. 2a, inset) although partially cut off by the filter. These early-time spectral features are explored in more depth with finer time resolution in the next section. We note that a flat or slightly rising $g_{lum}$ at early times is also apparent in the work of Hananel[22], though in that study the earliest time points after excitation were not sampled. Overall, both in the steady state and long timescale time-resolved measurements our setup gives results consistent with previous literature for Eu[(+)-facam]$_3$ in DMSO.

*Nanosecond gating of Eu[(+)-facam]$_3$*

Having established consistency with available literature in the steady-state and long timescale time-resolved CPL of Eu[(+)-facam]$_3$ in DMSO, we turn to time-resolved CPL on short timescales with 2-5 ns gating in the next step (Fig. 3). To the best of our knowledge, this is the first report of such data.

The time-resolved total luminescence spectra (Fig. 3a) reveal a number of additional spectral features at early times besides the typical $^5D_0 \rightarrow {}^7F_{1/2}$ transitions. The most prominent one is a broad unstructured peak around 435 nm (exact peak position may be affected by transmission/sensitivity at the blue edge) and a series of narrower peaks in the 520-570 nm range. For the assignment of these additional features in the early-time spectra of Eu[(+)-facam]$_3$, it is useful to briefly discuss the photophysical processes occurring in europium(III) complexes[32]. The $Eu^{3+}$ ion shows a multitude of well-studied spectral lines arising from $^5D_{J1} \rightarrow {}^7F_{J2}$ transitions[34,37]. These transitions are weakly absorbing (due to being weak magnetic dipole transitions or Laporte forbidden electric dipole transitions[38,39]), but strongly absorbing ligands with the right energy levels can be used as antennae that sensitize the emissive f-f excited states in the lanthanide ion[40].

Initially, a ligand-centered state ($^1LC$) is excited that can transfer the energy to the $Eu^{3+}$ ion after intersystem crossing (ISC) to a $^3LC$ triplet state[41]. Incomplete ISC or energy transfer can result in luminescence from the ligand-centered states[42]. As energy transfer occurs preferentially *via* the $^5D_1$ level[43] per the respective selection rules, emission from $^5D_1$ and even the higher-energy states $^5D_2$ and $^5D_3$ is occasionally observed, typically with a much shorter decay time than the main $^5D_0$ emission[32].

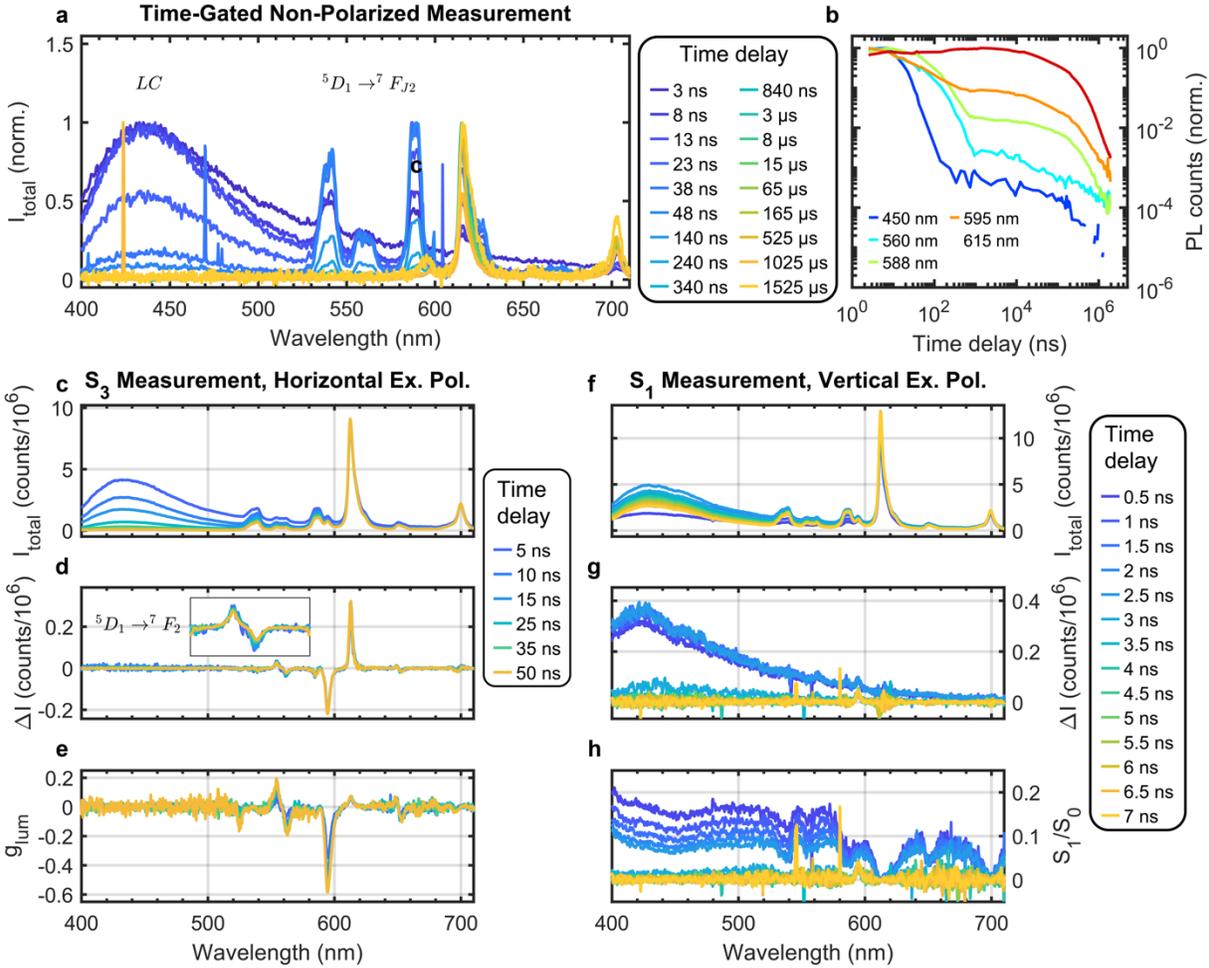

**Fig. 3 | Nanosecond time-resolved non-polarized luminescence, circular anisotropy and photoselection-induced linear anisotropy of Eu[(+)-facam]$_3$ in DMSO solution. a,** Non-polarization sensitive luminescence spectra (excitation 343 nm, 200 fs, 500 Hz) with varying time bins, normalized by maximum value to highlight spectral changes. Key early-time spectral features are labelled as the broad peak around 400-500 nm (LC), and a series of peaks 520-600 nm ($^5D_1 \rightarrow {}^7F_{J2}$). **b,** Kinetic traces of the non-polarization sensitive luminescence signal, at wavelengths corresponding to key features in spectra. **c-e**, Time-resolved CPL spectra (excitation 343 nm, 200 fs, 50 kHz) with 5 ns bins using horizontal excitation polarization to avoid photoselection. The transition $^5D_1 \rightarrow {}^7F_2$ at 560 nm is shown magnified in an inset and labelled. **f-h,** Time-resolved S$_1$ linear polarization measurement (excitation 343 nm, 200 fs, 50 kHz) with 2 ns bins and 0.5 ns steps, using vertical excitation polarization to intentionally induce photoselection (note that due to the time step and bin width, early-time bins fall within instrument response).

The features observed for Eu[(+)-facam]$_3$ are consistent with such a mechanism, showing features consistent with LC, $^5D_1 \rightarrow {}^7F_{J2}$ and $^5D_0 \rightarrow {}^7F_{J2}$ with progressively longer lifetimes. Kinetics at various wavelengths are collected in Fig. 3b, and multi-exponential fits are collected in the SI (Fig. S7). Due to the large number of timescales and overlapping features (LC and $^5D_1\rightarrow{}^7F_{J2}$ luminescence have different lifetimes, and $^5D_0\rightarrow{}^7F_{J2}$ luminescence has previously been reported as biexponential[22]) extracting robust lifetimes is difficult. Estimated timescales are 13 ns for the broad, unstructured LC emission and 130 ns for the $^5D_1\rightarrow{}^7F_{J2}$ emission lines. Both are very short compared to the $^5D_0\rightarrow{}^7F_{J2}$ lines (lifetimes on the order of 100 μs) which comprise the main bands dominating steady-state measurements.

To determine whether these short-lived features exhibit CPL, we have performed a time-resolved CPL measurement with gate steps and a gate width of 5 ns (Fig. 3c-e). For minimizing photoselection effects, a horizontal excitation polarization was used. To achieve sufficient signal-to-noise for CPL, a 50 kHz laser repetition rate was necessary, and therefore some roll-over emission is present (data with 50 ns gate steps and 500 Hz repetition rate to avoid roll-over is shown in the SI, Fig. S10). Both LC and $^5D_1\rightarrow{}^7F_{J2}$ luminescence are clearly observed. The ligand-centered transition does not appear to be significantly CPL-active. Even if some dissymmetry were present, we might expect this to be far smaller than that of $Eu^{3+}$ transitions and within the noise level of this measurement. In contrast, some $^5D_1\rightarrow{}^7F_{J2}$ transitions possess significant luminescence dissymmetry. In particular, the $^5D_1\rightarrow{}^7F_2$ transition around 560 nm shows strong bisignate CPL with $|g_{lum}| \approx 0.2$ as might be expected since it is a magnetic dipole transition just like the strongly dissymmetric $^5D_0\rightarrow{}^7F_1$ transition at 595 nm[38].

To rule out any contribution from linear polarization artifacts on short timescales, we intentionally induced photoselection by exciting Eu[facam]$_3$ with vertically polarized light and performed an $S_1$ linear polarization measurement with 2 ns oversampled (partially overlapping) time bins (Fig. 3f-h). Linear polarization is present in the LC emission band, which almost completely disappears within the instrument response (around 2 ns). Although $Eu^{3+}$ emission is observed, these features exhibit far less linear polarization, as might be expected since the emissive states are populated via energy transfer and not direct excitation. Our results are consistent with a strong photoselection of ligand dipoles followed by rapid reorientation on a timescale within the instrument response (2 ns). Even when intentionally maximized, significant linear polarization is not present on timescales used for the CPL measurements in Fig. 3c-e, increasing our confidence in those findings.

To summarize, TRCPL with nanosecond time resolution allows us to observe several short-lived electronic transitions in a europium(III) complex and evaluate their CPL activity for the first time. Hence, the method provides a handle to uncover signals which are otherwise suppressed in steady-state and microsecond time-resolved measurements but may be crucial for understanding photophysical relaxation pathways.

*Chiral organic delayed fluorescence emitter*

While strongly CPL-active chiral lanthanide complexes provide a convenient benchmark, studies on chiral emitters often involve materials with much weaker dissymmetry and faster luminescence decay. For example, purely organic small molecules in solution rarely exceed dissymmetry factors of $10^{-2}$ even in best-performing materials[26,44]. Hence, we demonstrate the broad applicability and sensitivity of our TRCPL setup using a CPL-active organic molecule. To introduce multiple timescales of emission, we have selected a chiral organic dye (*R/S*)-BINOL-phthalonitrile-tBuCz ((*R/S*)-BPC, Fig. 4) which shows thermally activated delayed fluorescence (TADF) and for which CPL has been previously characterized in the steady state[45].

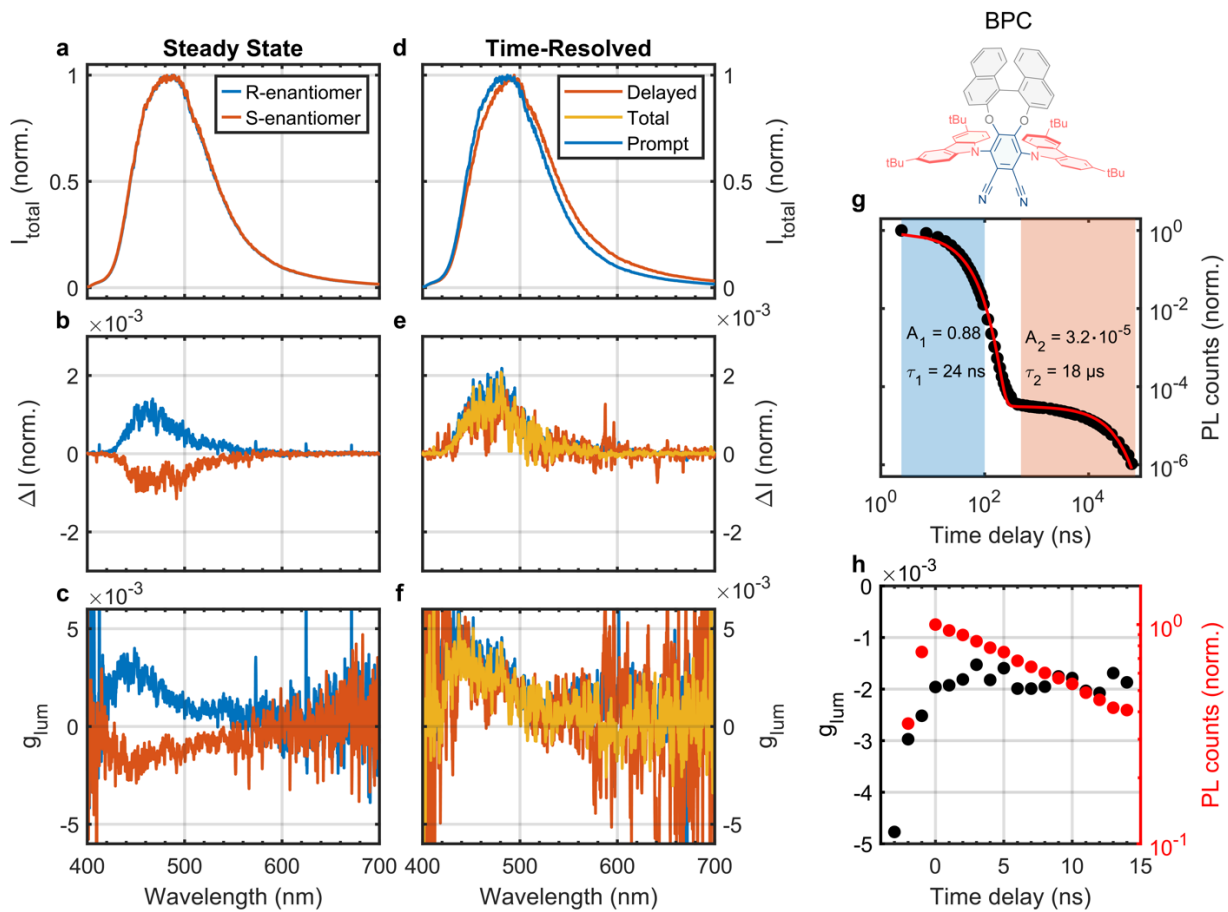

**Fig. 4 | Steady-state and time-resolved CPL experiments of a chiral TADF-active dye BPH (structure top right) in toluene solution. a-c**, Steady-state CPL spectra of (*R*)- and (*S*)-enantiomers (excitation 343 nm, 200 fs, 12.5 kHz). **d-f**, Time-resolved CPL spectra of the (*R*)-enantiomer. 'Prompt' refers to the first 100 ns, 'Delayed' to approximately 500 ns-80 μs, and 'Total' to a gate covering the complete emission process (0-80 μs). **g**, Kinetic trace of the total intensity decay integrated over the full spectrum, with highlighted regions showing the 'Prompt' and 'Delayed' time regions and parameters of a biexponential fit (red trace). **h**, Kinetic traces of the dissymmetry factor (average from 445-455 nm) and total intensity decay (integrated over 400-450 nm) with 3 ns time bins and 1 ns steps for the *S*-enantiomer (50 kHz repetition rate).

Steady-state CPL spectra of the two enantiomers in toluene solution show the expected mirror-image CPL (Fig. 4a-c) with $g_{lum}$ values of $+1.8\times10^{-3}$ (*R*) and $-1.3\times10^{-3}$ (*S*) at the peak wavelength matching those reported previously ($|g_{lum}| = 1.6\times10^{-3}$)[45]. To introduce time resolution, we have performed two experiments: first, we have separately gated the prompt, delayed and total emission components of R-BPC (Fig. 4d-f). The kinetic trace of the total emission (Fig. 4g) clearly displays a biexponential decay process. Compared to the delayed component, prompt emission is approximately 1000 times shorter-lived (24 ns *vs*. 18 μs), but over 10,000 times more intense. Consequently, the total emission spectrum is dominated by the prompt component, such that the spectra of prompt and total emission completely overlap (Fig. 4d). There appears to be a slight unexpected spectral shift between the prompt and delayed components. The emitting state in TADF is expected to be the same charge transfer ($^1$CT) state in both time regimes. In principle, the shift could arise from delayed fluorescence overlapping with another long-lived species such as phosphorescence from a $^3$CT state or a localized state ($^3$LE). Regardless of this slight spectral shift, the dissymmetry of the delayed component is not significantly affected, and the CPL/$g_{lum}$ spectra (Fig. 4e-f) are indistinguishable within noise for the different components. This not only confirms that the TRCPL instrument is sensitive enough to accurately quantify $g_{lum}$ on the order of $10^{-3}$, but it can also do so reliably for a temporally separated emission component comprising less than 1% of the total emission intensity.

Finally, we demonstrate that the setup can also collect clean kinetic traces of weak CPL signals with fine time steps of S-BPC (Fig. 4h). The first few bins show a rise caused by the instrument response. As convolution and division are not commutative, dissymmetry factors in this region should not be interpreted[46]. The value of $g_{lum}$ remains constant over the remaining part of the measured 15 ns time range, demonstrating the ability of our setup to track $g_{lum}$ values on the order of $10^{-3}$ on the nanosecond timescale. Temporal characterization of CPL in this material, which combines ns-scale and μs-scale decays with $g_{lum}$ of order $10^{-3}$, would not be feasible with pre-existing techniques.

*Polarization artefacts and relaxation in an achiral dye*

With a last experimental data set on the achiral standard dye Rhodamine B (Fig. 5) we demonstrate the low-noise zero baseline of our setup when photoselection is minimized, and the time-evolution of various apparent polarization components when photoselection is induced on purpose.

First, we measured the steady-state luminescence spectra of Rhodamine B in water as a relatively low-viscosity solvent with different excitation polarizations (Fig. 5a-c). Specifically, we used horizontal excitation polarization to minimize and vertical excitation polarization to intentionally maximize photoselection effects. The total emission spectrum is not impacted by the excitation polarization (Fig. 5a) and horizontally polarized excitation yields a flat baseline about 0 with noise at or below $10^{-3}$ for all polarization components $S_1$, $S_2$ and $S_3$, as expected for an achiral small molecule in solution (Fig. 5b-c, dashed lines).

When vertically polarized excitation is used, non-zero values for all polarization components appear (Fig. 5b-c, solid lines).

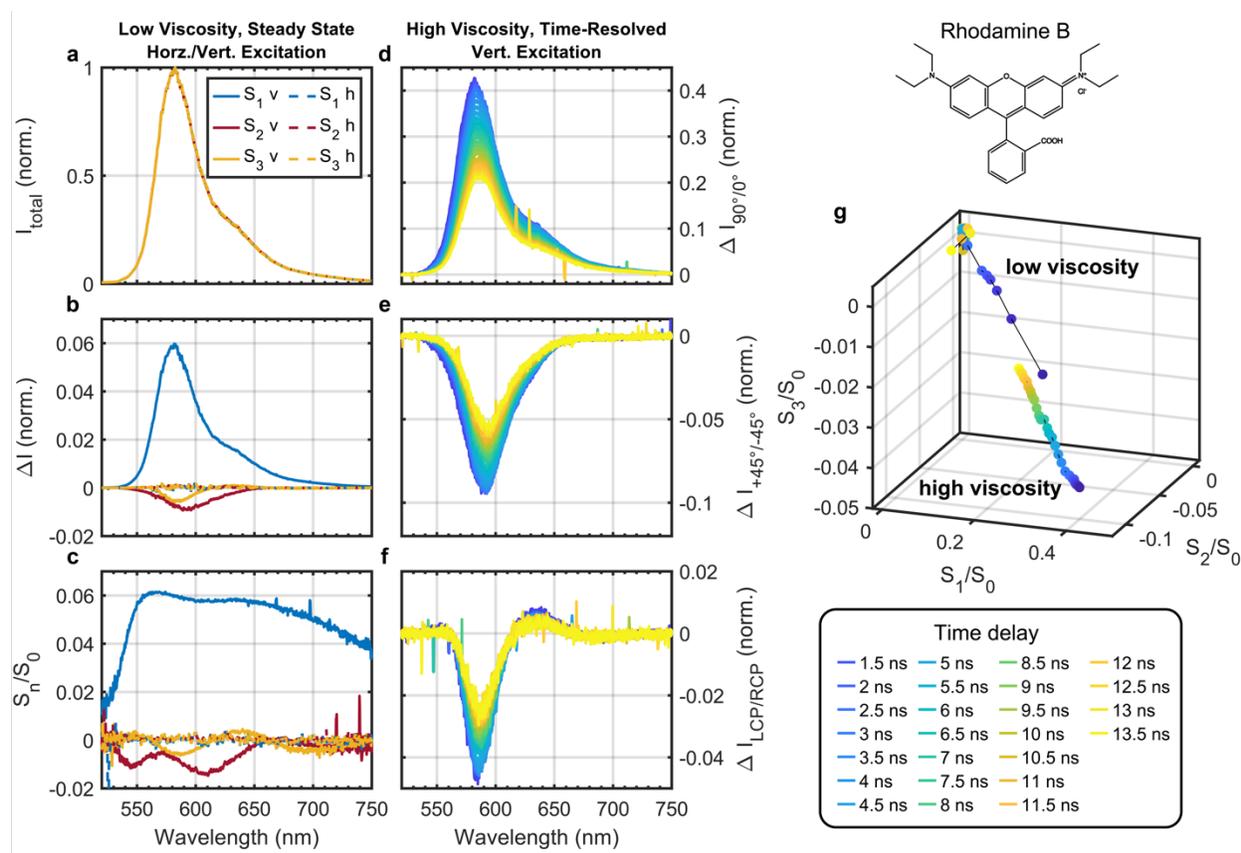

**Fig. 5 | Broadband steady-state and time-resolved full Stokes vector spectroscopy of a standard achiral dye (rhodamine B, structure top right) in aqueous solutions with low and high viscosity. a-c**, Steady state Stokes vector measurements (excitation 515 nm, 200 fs, 50 kHz) in a low-viscosity environment with horizontal and vertical excitation polarizations to inhibit and induce photoselection effects, respectively. Vertical excitation polarizations correspond to solid lines, horizontal excitation polarizations to dashed lines. We point out that all spectra overlap in **a**, and in **b** and **c** the dashed lines are all flat about 0. **d-f**, Time resolved intensity differences (excitation 515 nm, 200 fs, 50 kHz) over the Stokes polarization basis (normalized to total intensity maximum) in a high-viscosity environment with vertical excitation polarization to induce photoselection. **g**, Time evolution of all three Stokes components (averaged over a 10 nm range about the emission peak) in low-viscosity and high-viscosity environments with vertical excitation polarization (2 ns time bins, 0.5 ns time steps).

Expectedly, the $S_1$ component (horizontal-vertical linear polarization) shows the greatest response (the non-zero $S_2$ component at 45° and 135° linear polarization might stem from a slight tilt of the excitation beam, so the excitation polarization is not perfectly vertical along the detection axis). Importantly, we were able to detect a significant non-zero $S_3$ component for an achiral sample, *i.e.*, a CPL artifact. Such artifacts induced by linearly polarized emission components are a major challenge in accurate CPL measurements[19,29] across setup designs[27]. They are usually attributed to imperfections in the optical components, such as residual static birefringence and circular dichroism[19,27].

For the purposes of our time-resolved CPL measurements, it is important to understand how the temporal evolution of linear polarization components is translated to CPL artifacts over time. This requires the presence of time-evolving linear polarization on sufficiently long timescales to be properly characterized by the instrument[47].

In solution, photoselection effects decay over time as the excited molecules rotate and randomize the orientation of their transition dipole moments. While this is a straightforward way to achieve time-evolving linear polarization of emission, the rotational relaxation of fluorophores in many common solvents is faster than the instrument response of our setup (~2 ns) but can be increased by solvent viscosity[47]. As the emission properties of Rhodamine B are somewhat environment-dependent[48], rather than selecting completely different solvents of low and high viscosity we have opted to keep water as solvent and add sucrose to increase the viscosity.

In the high-viscosity environment with maximized photoselection, a very large $S_1$ component is present immediately after excitation (Fig. 5d) and decays over time. Smaller $S_2$ and $S_3$ components are also measured and depolarize over time (Fig. 5e-f). Their spectral shape and magnitude relative to $S_1$ is consistent with the steady-state measurements in a low-viscosity medium (Fig. 5b).

The time evolution of the Stokes parameters about the emission maximum in low and high-viscosity solutions is plotted in Fig. 5g. For the low-viscosity solution, depolarization occurs almost completely within the instrument response and the Stokes parameters are subsequently clustered around 0. In the high-viscosity solution, emission remains partially polarized after 10 ns, tracing an approximately straight line through the $S_1/S_2/S_3$ space (Fig. 5g). The $S_3$ component, representing a CPL artifact, is therefore proportional to the real $S_1$ component (though considerably smaller in magnitude)[27].

Our results demonstrate the necessity of controlling for linear polarization-induced artifacts in CPL, for which the ability to measure linear polarization components is important. Even in isotropic samples linear anisotropy may be induced by photoselection, and for high-viscosity solutions or solid-state samples this may persist beyond instrument response. Even in such cases, the impact of photoselection effects can be minimized by appropriate measurement parameters.

## Conclusion and discussion

We developed time-resolved broadband full-Stokes-vector luminescence spectroscopy as a versatile method for the polarization-resolved investigation of excited state dynamics. Our design establishes broadband nanosecond time resolution at millisecond range with a sensitivity noise floor of $10^{-4}$.

We then demonstrated the use of this setup in probing various timescales and degrees of CPL activity.

First, we validated our setup using the CPL standard Eu[(+)-facam]$_3$, for which we reproduced previously reported steady-state and µs-scale time-resolved spectra. Leveraging the nanosecond time-resolution of our instrument, we revealed the polarization-resolved luminescence dynamics of ligand-centered excited states and high-energy f-f transitions in $Eu^{3+}$ for the first time.

The high sensitivity of our method allowed us to track the temporal evolution (early-time-ns to late-time-µs) of even weak CPL signals in a chiral TADF emitter with dissymmetry factors on the order of $10^{-3}$, marking the first CPL dynamics report of such a material.

Lastly, we mapped out the temporal dynamics of polarization components and artifacts covering the full Stokes vector by deliberately introducing photoselection in an achiral dye. We unveil that CPL artifacts may also show time dynamics for certain timescales and experimental parameters, but also show that such artifacts can be effectively mitigated with appropriate experimental design. To facilitate adoption by the wider community we share the full setup design, along with an open-source software package used for measurement automation. In addition, we provide a compendium of practical considerations to ensure accurate quantification of weak polarization signals with special attention to non-obvious error sources such as beam deflection and pixel-to-pixel variations in detector sensitivity. Such factors open new avenues to further push the limits of the sensitivity and time resolution in the future, *e.g.*, by closer consideration of beam drift and detector sensitivity interactions as well as employing optical gating for covering ultrafast timescales.

Overall, broadband time-resolved full-Stokes luminescence spectroscopy dramatically expands the sensitivity, timescale and scope of accessible polarization information compared to the state-of-the-art. Importantly, all of this is achieved while retaining a straightforward spectrometer design based off stock components. We hope that our work will fuel the development of next-generation high-performance optical materials by revealing their underlying dynamics with unprecedented detail.

*Acknowledgments*

The authors acknowledge Winald R. Kitzmann for helpful discussions and the staff at the Rowland Institute at Harvard, in particular R. Christopher Stokes, for technical assistance.

*Funding*

The authors are grateful to the Rowland Institute at Harvard, the Studienstiftung des deutschen Volkes, and EPFL for providing funding.

*Competing interests*

A patent concerning the transient broadband sensitive full-Stokes vector setup reported here has been recently filed.

*Data availability*

All data are available in the main text, the supplementary materials, or the online repository.

# Supplementary Information

**Broadband Transient Full-Stokes Luminescence Spectroscopy with High Sensitivity**

Antti-Pekka M. Reponen[1], Marcel Mattes[1,2], Zachary A. VanOrman[1,2], Lilian Estaque[3], Grégory Pieters[3], and Sascha Feldmann[1,2]*

[1]Rowland Institute, Harvard University, Cambridge, MA, USA

[2]Institute of Chemical Sciences and Engineering, École Polytechnique Fédérale de Lausanne, Lausanne, Switzerland

[3]Université Paris-Saclay, CEA, INRAE, Département Médicaments et Technologies pour la Santé, Gif-sur-Yvette, France

*Email: sascha.feldmann@epfl.ch

# 1: Materials and Methods

## 1.1: List of Setup Components

Pulsed excitation source: Light Conversion PHAROS (1030 nm, 180 fs, 50 kHz with pulse picker) (Light Conversion HIRO harmonic unit output at 515 nm/343 nm used)

CW excitation source: ThorLabs DL5146-101S 405 nm laser diode

Wollaston prism: ThorLabs WPQ10 1º beam separation, uncoated quartz, 400-2000 nm

Quarter-wave plate: ThorLabs SAQWP05M-700 superachromatic 325-1100 nm

Half-wave plate (detection/excitation): ThorLabs SAHWP05M-700 superachromatic 310-1100 nm

Linear polarizer (excitation): ThorLabs WP25M-UB

Waveplate motorized housing: ThorLabs PRM1/MZ8

Waveplate housing motor controller: ThorLabs KDC101

L2: Focusing lens (excitation): ThorLabs LA4579 UV fused silica, uncoated, f = 301.1 mm plano-convex

L3: Collimating lens (detection): ThorLabs LA4380 UV fused silica, uncoated, f = 100.3 mm plano-convex

L4: Focusing lens (detection): ThorLabs AC508-100-A

F: 375/450/530 nm longpass filter (depending on excitation source)

Spectrograph: Andor Kymera 328i

Grating: SR-GRT-0150-0500 150 l/mm 500 nm blaze reflecive grating

Detector: Andor DH340T-18U-74 gated iCCD camera

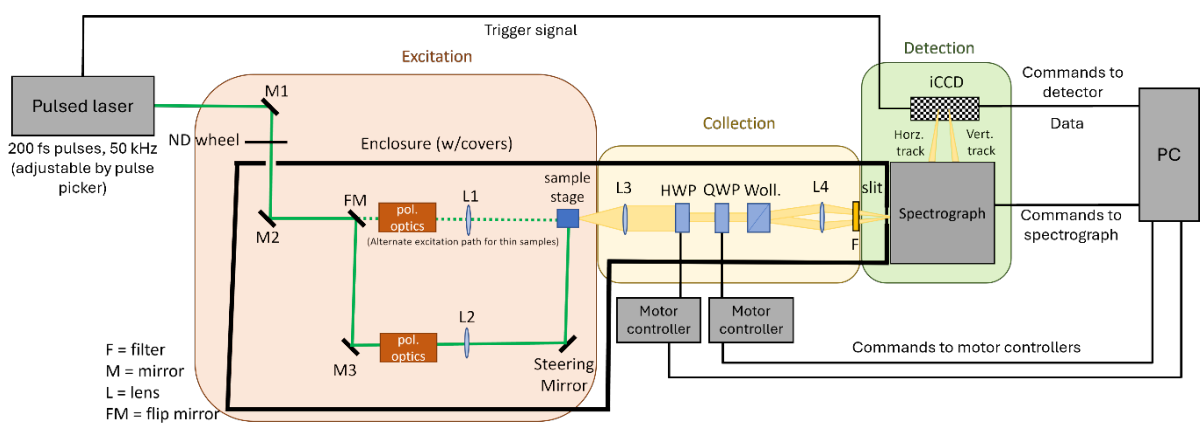

*Figure S1: Setup schematic showing the main components, beam paths and electronic signals in the instrument.*

## 1.2: Sample Preparation

Eu[(+)-facam]$_3$ was used as received from the supplier (Sigma-Aldrich). Rhodamine B was used as received from the supplier (Radiant Dyes). (R/S)-BINOL-phthalonitrile-tBuCz was synthesized according to previously described methods (*45*).

Solution concentrations were 0.5 mM (non-polarization resolved measurements) and 11 mM (polarization-resolved measurements) for Eu[(+)-facam]$_3$ in DMSO, 0.6 mM for (R/S)-BINOL-phthalonitrile-tBuCz in toluene, 37 μM for Rhodamine B in water, 33 μM for Rhodamine B in a water/sucrose mixture.

Solution samples were prepared using dry solvents in a nitrogen-filled glovebox, except aqueous solutions of Rhodamine B which were prepared under ambient conditions. Aqueous solutions were prepared using distilled water; for high-viscosity solutions, sucrose was dissolved in water near the solubility limit (approx.. 2 g/ml).

Solutions were placed in screw-top four-windowed quartz cuvettes with a square base 1 cm across. Air ingress to cuvettes was reduced by sealing with PTFE tape and parafilm within the glovebox, where appropriate.

## 1.3: Sample Excitation

For time-resolved measurements samples were optically excited using the output of a Light Conversion PHAROS laser (Yb:KGW lasing medium, 1030 nm, pulse energy 400 μJ, pulse width duration 200 fs, repetition rate of 50 kHz). The pump beam was generated from the seed in a harmonic generation unit (Light Conversion HIRO) *via* nonlinear crystals (beta-barium borate, lithium triborate) with residual fundamental removed by dichroic mirrors within the unit. Second and third harmonics can be generated, giving pump wavelengths of 515 nm or 343 nm respectively. Pump pulse energy at the sample was 10-70 nJ, with the pump focused down to a beam diameter of approximately 1 mm. The laser repetition rate is controllable by a pulse picker, and repetition rates in the range 500 Hz – 50 kHz were used (specified where data is presented).

For continuous wave excitation at 405 nm, a laser diode (ThorLabs DL5146-101S mounted in a ThorLabs LDM9T temperature-controlled mount) was used, with a constant output power of 5-50 mW.

## 1.4: Data Collection/Processing

The CCD sensor has 2048x512 pixels, enabling simultaneous recording of multiple tracks. Horizontal and vertical polarization components, which are spatially separated by a Wollaston prism (shown on the sensor in Figure S2), can therefore be simultaneously recorded. Vertical pixel binning is used to produce two effective vertical pixels for each wavelength pixel, giving $I_h(\lambda)$ and $I_v(\lambda)$ for the horizontal and vertical channels respectively.

For time-resolved measurements, the quantities recorded during a single acquisition are $I_h(\lambda, t)$ and $I_v(\lambda, t)$ where $t$ is defined by the gate pulse applied (illustrated in Figure S3). Time series are built up by repeating the measurement with modified gate delays/widths.

Data collection to the level of $I_h(\lambda, t)$ and $I_v(\lambda, t)$ occurs entirely within the Andor Solis T (time-resolved) software platform (Oxford Instruments Andor Ltd).

Error cancellation and changing the Stokes polarization component measured requires repeating the measurement with rotated waveplates. Subsequent data processing to obtain error-corrected polarization spectra and related quantities as well as processing for plotting and lifetime fitting was carried out with MATLAB (*50*) scripts. Waveplate operation and calculations for data processing are described main body and in following SI section.

Where appropriate, the transmission curve of longpass filters used was measured and corrected for in data processing. Transmission characters of other optical components were not corrected for.

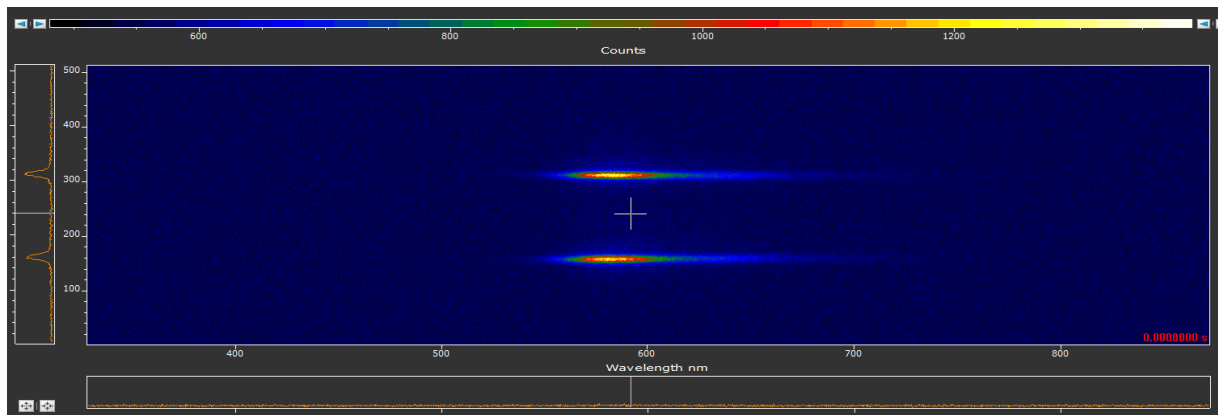

*Figure S2: Full-sensor image showing the two orthogonal polarization tracks. For data acquisition, the top half and bottom half are vertically binned for faster readout and less readout noise, resulting in two effective vertical pixels.*

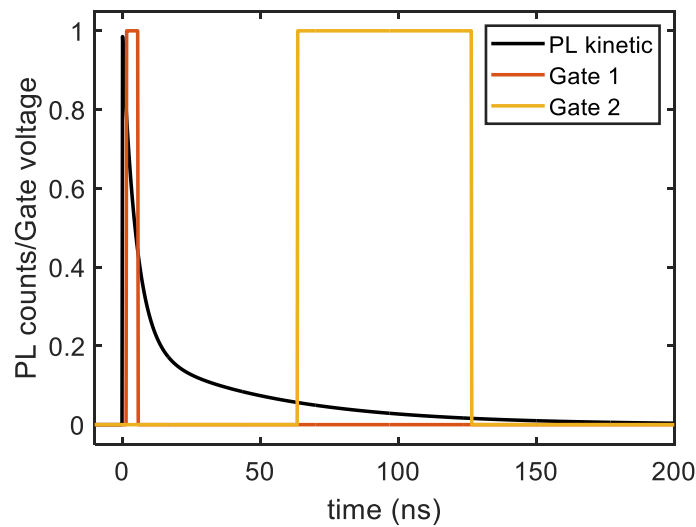

*Figure S3: Illustration of time-gating. The sample is excited by a laser pulse arriving at t=0, with the laser sync signal triggering a gate pulse on the detector with a user-defined delay and width. The detector only collects light when the gate pulse is on, resulting in collection of time-resolved spectra.*

## 2: Supplementary Text

### 2.1: Waveplate Interactions and Error Cancellation

Polarization tracks are separated (vertically, for incoupling through a vertical slit) from a single beam by a Wollaston prism which remains fixed, therefore one track will always correspond to a vertical polarization and the other to a horizontal polarization. The basic idea is to use waveplates to make the polarization components of interest (0º/90º, +45º/-45º, LCP/RCP) fall completely to the horizontal or vertical tracks, and then use another waveplate orientation to swap the components on the tracks.

This swapping of tracks, combined with measuring both tracks at two such waveplate orientations, allows for cancelling out the bulk of errors caused by any transmission inequalities between the two beampaths, as described by Baguenard *et al.*(*27*). Further, the simultaneous recording of both tracks cancels out the bulk of time-instability errors, arising from e.g. drift in excitation intensity. This is illustrated in Figure S4.

After the Wollaston prism, the channels are horizontally and vertically polarized (0º/90º respectively), corresponding to the $S_1$ polarization axes. To investigate $S_2$ (linear ±45º) and $S_3$ (LCP/RCP) polarization components, waveplates are used to transform these polarization components to 0º/90º linear polarizations. For $S_2$, suitable polarization transformation is achieved by a half-wave plate (HWP) at 22.5º ± 45º, and for $S_3$ with a quarter-wave plate (QWP) at 45º ± 90º, where the waveplate orientations are defined about the fast axis relative to the table plane. Any two adjacent orientations will swap the polarization component imaged on a given track; for example, if the RCP component is measured at the horizontal track for a 45º QWP orientation, at a 135º QWP orientation the RCP component is measured at the vertical track.

In principle, the setup would only require a HWP for $S_1/S_2$ measurement and only a QWP for $S_3$ measurement. For a faster, automatable full-Stokes measurement, it is advantageous to have both waveplates in place for all measurements. A quick illustration of how the various Stokes components produce intensity differences between the tracks in the ideal case (perfect waveplates) with both a HWP and QWP present is presented in Figure 1c of the main body. Importantly, for CPL measurements the HWP orientation in principle does not matter, but in all orientations the HWP will act to transform LCP to RCP and vice versa which must be accounted for in processing.

In all cases, the waveplate angles for measuring a given $S_n$ component are such that the other two components are split evenly across the two channels and therefore are not measured as a false polarization signal. Of course, imperfections in real-life optics are well known to produce polarization artifacts, including the well-known false CPL signals when significant linear polarization is present (*19*). Therefore, as is good practice in polarization measurements, care must be taken when interpreting small polarization components in the presence of other, larger polarization components.

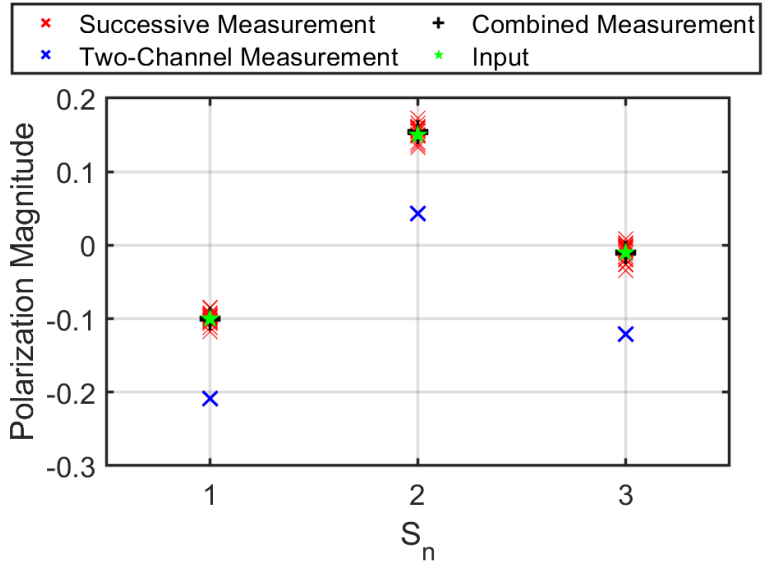

*Figure S4: Simulated example illustrating how combining successive measurements by waveplate rotation and simultaneous measurements by two-channel detection can accurately recover all Stokes components even with substantial spatial and temporal error sources. Here, track transmission mismatch is set to 20% and a random time drift set to 5% of the incident light intensity for this case.*

## 2.2: Measured and Calculated Quantities

As described above, two tracks are simultaneously recorded at a given orientation of the HWP and QWP. Accordingly, we will call the recorded intensities $I_{v,QWP\theta,HWP\theta}$ and $I_{h,QWP\theta,HWP\theta}$.

An $S_3$ measurement would then seek to find $I_{LCP}$ and $I_{RCP}$, recorded as

$$I_{LCP} = I_{h,QWP45°,HWP0°} + I_{v,QWP135°,HWP0°}$$

$$I_{RCP} = I_{v,QWP45°,HWP0°} + I_{h,QWP135°,HWP0°}$$

From which we calculate the quantities

$$\Delta I = I_{LCP} - I_{RCP}$$

$$I_{total} = I_{LCP} + I_{RCP}$$

Whereas for an $S_1$ measurement we have

$$I_{0°} = I_{h,QWP0°,HWP0°} + I_{v,QWP0°,HWP45°}$$

$$I_{90°} = I_{v,QWP0°,HWP0°} + I_{h,QWP0°,HWP45°}$$

$$\Delta I = I_{0°} - I_{90°}$$

And for an $S_2$ measurement

$$I_{+45°} = I_{h,QWP0°,HWP22.5°} + I_{v,QWP0°,HWP67.5°}$$

$$I_{-45°} = I_{v,QWP0°,HWP°} + I_{h,QWP0°,HWP67.5°}$$

$$\Delta I = I_{+45°} - I_{-45°}$$

## 2.3: Stokes Components vs. Anisotropy/Dissymmetry Values

The Stokes components are defined as

$$S_0 = I_{0°} + I_{90°} = I_{+45°} + I_{-45°} = I_{LCP} + I_{RCP}$$

$$S_1 = I_{0°} - I_{90°}$$

$$S_2 = I_{+45°} - I_{-45°}$$

$$S_3 = I_{LCP} - I_{RCP}$$

Since the absolute count numbers are generally not of interest, for convenience we may scale the data such that $\max(S_0) = 1$. Alternatively, we can calculate the relative quantity $S_{1/2/3}/S_0$ in cases where the polarized proportion at each wavelength is to be presented.

Note that the quantities $S_{1/2/3}$ (even if normalized by $S_0$) which describe the polarization state of light are similar but not exactly the same as the commonly used quantities for describing material properties, these being the linear polarization anisotropy ($r$) and circular polarization dissymmetry of the luminescence ($g_{lum}$) which are instead defined as (*46*)

$$r = \frac{I_\parallel - I_\perp}{I_\parallel + 2I_\perp}$$

where $I_\parallel$ and $I_\perp$ are parallel and perpendicular linearly polarized intensities, and

$$g_{lum} = \frac{I_{LCP} - I_{RCP}}{I_{avg}} = \frac{I_{LCP} - I_{RCP}}{\frac{1}{2}(I_{LCP} + I_{RCP})}$$

Therefore, some care is required when comparing to other work where different conventions may be used. Our main literature comparisons relate to circular dissymmetries, and therefore where only CPL is measured we report $g_{lum}$ to make those comparisons simpler. When multiple polarization components are presented we use the Stokes parameters for internal consistency.

## 2.4: Time Resolution, Range and Repetition Rate

Electronic gating in the current setup has a minimum gate width of approximately 2 ns, which is also the overall limit of time resolution (as the gate delay accuracy and laser pulse width are far smaller).

Maximum time range is practically limited by the laser repetition rate, as very long gate pulse values are achievable. In our setup, the laser operates at 50 kHz and can be pulse-picked for a lower frequency operation at the same per-pulse fluence (and lower time-averaged power).

Lower repetition rates usually lead to slower data acquisition. In our experiments, the practical limit to perform polarization measurements is at approximately 500 Hz, corresponding to a time range of 2 ms; extending this would be reasonably straightforward by increasing per-pulse power at the sample, in principle.

Overall, accessible time bins range from approximately 2 ns – 2 ms.

For time-averaged values, corresponding to steady-state measurements, a bin size spanning the entire time period between excitation pulses (or at least the time period during which luminescence is present) can be used with a pulsed excitation, or the setup can be switched to a continuous wave (CW) excitation source, in which case only a time-averaged value is measured regardless of detector time gating.

**2.5: Photoselection From Polarized Excitation**

Photoselection arises from preferential excitation of dipoles along the electric field (polarization) axis of excitation light. It is important to consider in CPL measurements due to their sensitivity to linear anisotropy-induced artifacts, with photoselection being a common source of linear anisotropy even in otherwise isotropic samples as well described by Blok and Dekkers (*29*).

Photoselection can never be truly eliminated, as the propagating light beam contains no electric field component parallel to its propagation direction for any polarization (including unpolarized light). Dipoles along this axis can therefore not be excited.

However, preferential orientation of dipoles along an axis parallel to the light collection axis will result in uniformly (un)polarized light at the collector. This is illustrated in the cartoon below, and is the situation when the excitation polarization is horizontal and the collection axis is orthogonal to the excitation axis.

On the other hand, using vertical polarization in the same geometry results in maximally polarized light along the collection axis.

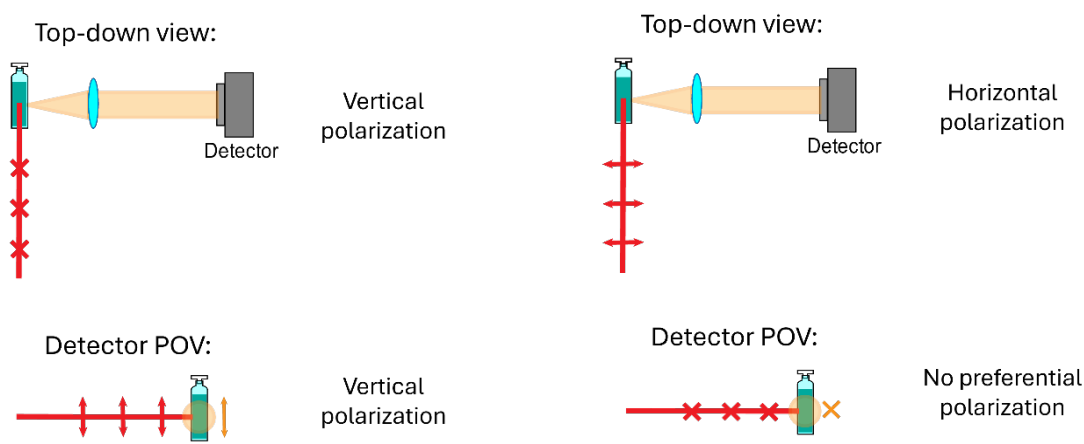

## 2.6: Signal-to-Noise and Acquisition Time in Time-Resolved CPL

We briefly explain why TCSPC-based approaches, such as described by Schauerte *et al.*(*18*) face issues with impractically long collection times.

At the most fundamental level, TCSPC signal acquisition rate is limited by the need to avoid photon pile-up effects. Noise in single photon counting measurements is well-described by Poisson statistics, where the signal-to-noise ratio (S/N) scales as $\sqrt{N}$ for $N$ photons counted. For noise levels on the order of $10^{-4}$ (desirable for accurate measurement of typical $g_{lum}$ of about $10^{-3}$), this requires counting $10^8$ photons. Typically, TCSPC measurements are carried out with photon detection rate not exceeding $0.05f$ where $f$ is the sync signal frequency (i.e. excitation source repetition rate) to avoid distorting kinetics through photon pile-up. It follows that achieving the $10^{-4}$ noise level *via* TCSPC requires measuring for $\frac{2\times 10^9}{f}$ seconds, practically necessitating MHz repetition rate excitation; even for a 1 MHz excitation, this gives a collection time of over half an hour. The problem is made worse when one takes into account that the photons are further distributed over multiple time bins, with most of the photons falling into the earlier bins. Achieving high S/N in the later time bins is therefore prohibitively time-consuming. As TCSPC detectors are single-element detectors, building up a spectrum also necessitates stepping through wavelengths one at a time, multiplying the required measurement time.

To mitigate issues, excitation sources with repetition rates on the order of 10 MHz could be used, and multiple counting channels can be used. However, even at 1 MHz the excitation pulse spacing is only 1 μs, insufficient for many materials emitting via spin- or symmetry-forbidden transitions (e.g. chiral lanthanide complexes (*24*)).

## 2.7: Müller Calculus Representation of the Measurement

For convenience and clarity, basic principles of the measurements are shown here. A very thorough error analysis treatment for CPL measurements specifically (equivalent to $S_3$ measurements here) is found in the work of Baguenard *et al.*(*27*) which we do not aim to reproduce in this work.

We begin with the collecting lens collimating a portion of light emitted by the sample. This light has a polarization state represented by the Stokes vector

$$S_{in} = \begin{pmatrix} S_0 \\ S_1 \\ S_2 \\ S_3 \end{pmatrix}$$

The QWP and HWP have Müller matrices (when their fast axis is vertical)

$$M_{QWP} = \begin{pmatrix} 1 & 0 & 0 & 0 \\ 0 & 1 & 0 & 0 \\ 0 & 0 & 0 & -1 \\ 0 & 0 & 1 & 0 \end{pmatrix}$$

$$M_{HWP} = \begin{pmatrix} 1 & 0 & 0 & 0 \\ 0 & 1 & 0 & 0 \\ 0 & 0 & -1 & 0 \\ 0 & 0 & 0 & -1 \end{pmatrix}$$

For arbitrary waveplate orientations $\theta$, we use the rotation and inverse rotation matrixes

$$R(\theta) = \begin{pmatrix} 1 & 0 & 0 & 0 \\ 0 & \cos(2\theta) & \sin(2\theta) & 0 \\ 0 & -\sin(2\theta) & \cos(2\theta) & 0 \\ 0 & 0 & 0 & 1 \end{pmatrix}$$

$$R^{-1}(\theta) = \begin{pmatrix} 1 & 0 & 0 & 0 \\ 0 & \cos(2\theta) & -\sin(2\theta) & 0 \\ 0 & \sin(2\theta) & \cos(2\theta) & 0 \\ 0 & 0 & 0 & 1 \end{pmatrix}$$

And the Wollaston prism (which remains in a fixed orientation) can be represented by separating the beam to two channels passing through orthogonal linear polarizers

$$P_{horizontal} = \frac{1}{2}\begin{pmatrix} 1 & 1 & 0 & 0 \\ 1 & 1 & 0 & 0 \\ 0 & 0 & 0 & 0 \\ 0 & 0 & 0 & 0 \end{pmatrix}$$

$$P_{vertical} = \frac{1}{2}\begin{pmatrix} 1 & -1 & 0 & 0 \\ -1 & 1 & 0 & 0 \\ 0 & 0 & 0 & 0 \\ 0 & 0 & 0 & 0 \end{pmatrix}$$

The Stokes vectors of the two tracks at detection are then given by

$$S_{horizontal} = P_{horizontal} R^{-1}(\theta_{QWP}) M_{QWP} R(\theta_{QWP}) R^{-1}(\theta_{HWP}) M_{HWP} R(\theta_{HWP}) S_{in} = M_h S_{in}$$

$$S_{vertical} = P_{vertical} R^{-1}(\theta_{QWP}) M_{QWP} R(\theta_{QWP}) R^{-1}(\theta_{HWP}) M_{HWP} R(\theta_{HWP}) S_{in} = M_v S_{in}$$

Where we have collected the total system matrices for the two channels as $M_h$ and $M_v$ and the polarization of the luminescence is described by the Stokes vector $S_{in}$. Though the output at detection is polarized in each case (being separated to horizontal/vertical components by the Wollaston prism), the detector itself only measures the total intensity (the Stokes component $S_0$). This is recorded as $I_h$ and $I_v$ for the horizontal and vertical channels respectively.

(More properly the detection elements, especially the grating, have some degree of polarization sensitivity. However, as the Wollaston is fixed, this will act analogously to a fixed transmission imbalance between the channels.)

To begin with, for an $S_1$ measurement we would use $\theta_{QWP} = 0°$ with $\theta_{HWP} = 0°$ and $45°$

Then for $\theta_{HWP} = 0$

$$S_{horizontal} = \frac{1}{2}\begin{pmatrix} 1 & 1 & 0 & 0 \\ 1 & 1 & 0 & 0 \\ 0 & 0 & 0 & 0 \\ 0 & 0 & 0 & 0 \end{pmatrix}\begin{pmatrix} S_0 \\ S_1 \\ S_2 \\ S_3 \end{pmatrix} = \begin{pmatrix} \frac{1}{2}(S_0 + S_1) \\ \frac{1}{2}(S_0 + S_1) \\ 0 \\ 0 \end{pmatrix} \text{ giving } I_{h,0°} = \frac{1}{2}(S_0 + S_1)$$

$$S_{vertical} = \frac{1}{2}\begin{pmatrix} 1 & -1 & 0 & 0 \\ -1 & 1 & 0 & 0 \\ 0 & 0 & 0 & 0 \\ 0 & 0 & 0 & 0 \end{pmatrix}\begin{pmatrix} S_0 \\ S_1 \\ S_2 \\ S_3 \end{pmatrix} = \begin{pmatrix} \frac{1}{2}(S_0 - S_1) \\ -\frac{1}{2}(S_0 - S_1) \\ 0 \\ 0 \end{pmatrix} \text{ giving } I_{v,0°} = \frac{1}{2}(S_0 - S_1)$$

and for $\theta_{HWP} = 45°$

$$S_{horizontal} = \frac{1}{2}\begin{pmatrix} 1 & -1 & 0 & 0 \\ 1 & -1 & 0 & 0 \\ 0 & 0 & 0 & 0 \\ 0 & 0 & 0 & 0 \end{pmatrix}\begin{pmatrix} S_0 \\ S_1 \\ S_2 \\ S_3 \end{pmatrix} = \begin{pmatrix} \frac{1}{2}(S_0 - S_1) \\ \frac{1}{2}(S_0 - S_1) \\ 0 \\ 0 \end{pmatrix} \text{ giving } I_{h,45°} = \frac{1}{2}(S_0 - S_1)$$

$$S_{vertical} = \frac{1}{2}\begin{pmatrix} 1 & 1 & 0 & 0 \\ -1 & -1 & 0 & 0 \\ 0 & 0 & 0 & 0 \\ 0 & 0 & 0 & 0 \end{pmatrix}\begin{pmatrix} S_0 \\ S_1 \\ S_2 \\ S_3 \end{pmatrix} = \begin{pmatrix} \frac{1}{2}(S_0 + S_1) \\ -\frac{1}{2}(S_0 + S_1) \\ 0 \\ 0 \end{pmatrix} \text{ giving } I_{v,45°} = \frac{1}{2}(S_0 + S_1)$$

For an $S_2$ measurement we would use $\theta_{QWP} = 0$ and $\theta_{HWP} = 22.5°$ and $67.5°$

Then for $\theta_{HWP} = 22.5°$

$$S_{horizontal} = \frac{1}{2}\begin{pmatrix} 1 & 0 & 1 & 0 \\ 1 & 0 & 1 & 0 \\ 0 & 0 & 0 & 0 \\ 0 & 0 & 0 & 0 \end{pmatrix}\begin{pmatrix} S_0 \\ S_1 \\ S_2 \\ S_3 \end{pmatrix} = \begin{pmatrix} \frac{1}{2}(S_0 + S_2) \\ \frac{1}{2}(S_0 + S_2) \\ 0 \\ 0 \end{pmatrix} \text{ giving } I_{h,22.5°} = \frac{1}{2}(S_0 + S_2)$$

$$S_{vertical} = \frac{1}{2}\begin{pmatrix} 1 & 0 & -1 & 0 \\ -1 & 0 & 1 & 0 \\ 0 & 0 & 0 & 0 \\ 0 & 0 & 0 & 0 \end{pmatrix}\begin{pmatrix} S_0 \\ S_1 \\ S_2 \\ S_3 \end{pmatrix} = \begin{pmatrix} \frac{1}{2}(S_0 - S_2) \\ -\frac{1}{2}(S_0 - S_2) \\ 0 \\ 0 \end{pmatrix} \text{ giving } I_{v,22.5°} = \frac{1}{2}(S_0 - S_2)$$

and for $\theta_{HWP} = 67.5°$

$$S_{horizontal} = \frac{1}{2}\begin{pmatrix} 1 & 0 & -1 & 0 \\ 1 & 0 & -1 & 0 \\ 0 & 0 & 0 & 0 \\ 0 & 0 & 0 & 0 \end{pmatrix}\begin{pmatrix} S_0 \\ S_1 \\ S_2 \\ S_3 \end{pmatrix} = \begin{pmatrix} \frac{1}{2}(S_0 - S_2) \\ \frac{1}{2}(S_0 - S_2) \\ 0 \\ 0 \end{pmatrix} \text{ giving } I_{h,67.5°} = \frac{1}{2}(S_0 - S_2)$$

$$S_{vertical} = \frac{1}{2}\begin{pmatrix} 1 & 0 & 1 & 0 \\ -1 & 0 & -1 & 0 \\ 0 & 0 & 0 & 0 \\ 0 & 0 & 0 & 0 \end{pmatrix}\begin{pmatrix} S_0 \\ S_1 \\ S_2 \\ S_3 \end{pmatrix} = \begin{pmatrix} \frac{1}{2}(S_0 + S_2) \\ -\frac{1}{2}(S_0 + S_2) \\ 0 \\ 0 \end{pmatrix} \text{ giving } I_{v,67.5°} = \frac{1}{2}(S_0 + S_2)$$

Finally, for an $S_3$ measurement we would use $\theta_{QWP} = 45°$ and $135°$ with $\theta_{HWP} = 0$ (or any other HWP angle, as the HWP will invert the handedness of incident light regardless of its orientation, effectively inverting the sign of $S_3$).

Then for $\theta_{QWP} = 45°$

$$S_{horizontal} = \frac{1}{2}\begin{pmatrix} 1 & 0 & 0 & -1 \\ 1 & 0 & 0 & -1 \\ 0 & 0 & 0 & 0 \\ 0 & 0 & 0 & 0 \end{pmatrix}\begin{pmatrix} S_0 \\ S_1 \\ S_2 \\ S_3 \end{pmatrix} = \begin{pmatrix} \frac{1}{2}(S_0 - S_3) \\ \frac{1}{2}(S_0 - S_3) \\ 0 \\ 0 \end{pmatrix} \text{ giving } I_{h,45°} = \frac{1}{2}(S_0 - S_3)$$

$$S_{vertical} = \frac{1}{2}\begin{pmatrix} 1 & 0 & 0 & 1 \\ -1 & 0 & 0 & -1 \\ 0 & 0 & 0 & 0 \\ 0 & 0 & 0 & 0 \end{pmatrix}\begin{pmatrix} S_0 \\ S_1 \\ S_2 \\ S_3 \end{pmatrix} = \begin{pmatrix} \frac{1}{2}(S_0 + S_3) \\ -\frac{1}{2}(S_0 + S_3) \\ 0 \\ 0 \end{pmatrix} \text{ giving } I_{v,45°} = \frac{1}{2}(S_0 + S_3)$$

and for $\theta_{QWP} = 135°$

$$S_{horizontal} = \frac{1}{2}\begin{pmatrix} 1 & 0 & 0 & 1 \\ 1 & 0 & 0 & 1 \\ 0 & 0 & 0 & 0 \\ 0 & 0 & 0 & 0 \end{pmatrix}\begin{pmatrix} S_0 \\ S_1 \\ S_2 \\ S_3 \end{pmatrix} = \begin{pmatrix} \frac{1}{2}(S_0 + S_3) \\ \frac{1}{2}(S_0 + S_3) \\ 0 \\ 0 \end{pmatrix} \text{ giving } I_{h,135°} = \frac{1}{2}(S_0 + S_3)$$

$$S_{vertical} = \frac{1}{2}\begin{pmatrix} 1 & 0 & 0 & -1 \\ -1 & 0 & 0 & 1 \\ 0 & 0 & 0 & 0 \\ 0 & 0 & 0 & 0 \end{pmatrix}\begin{pmatrix} S_0 \\ S_1 \\ S_2 \\ S_3 \end{pmatrix} = \begin{pmatrix} \frac{1}{2}(S_0 - S_3) \\ -\frac{1}{2}(S_0 - S_3) \\ 0 \\ 0 \end{pmatrix} \text{ giving } I_{v,135°} = \frac{1}{2}(S_0 - S_3)$$

As can be seen, the outputs for $S_2$ and $S_3$ are analogous to results obtained for $S_1$, and therefore treatment for recovering the Stokes components is very similar in each case. Accordingly, we only explicitly show the process for $S_1$.

We ended up with the various measured intensities $I$:

$$I_{h,0°} = \frac{1}{2}(S_0 + S_1) \qquad I_{v,0°} = \frac{1}{2}(S_0 - S_1)$$

$$I_{h,45°} = \frac{1}{2}(S_0 - S_1) \qquad I_{v,45°} = \frac{1}{2}(S_0 + S_1)$$

From which we can straightforwardly acquire the $S_0$ or $S_1$ components.

As currently defined, having two waveplate orientations and spatial channels appears superfluous. However, we need to consider the effect of imperfections which result in intensity differences between channels for reasons besides the presence of real polarization components. The strongest effects are usually those proportional to total luminescence intensity, which can be roughly grouped under transmission imbalances and time drift-like effects.

**Transmission Imbalance**

A transmission imbalance between the horizontal and vertical channels can result either from imperfectly uniform beam paths or unequal polarization responses. We can represent this by adding transmission factors $T_h$ and $T_v$ (unknown numbers between 0 and 1) to the horizontal and vertical paths respectively:

$$S_{horizontal} = T_h M_h S_{in}$$

$$S_{vertical} = T_v M_v S_{in}$$

The outputs are then:

$$I_{h,0°} = \frac{1}{2}T_h(S_0 + S_1) \qquad I_{v,0°} = \frac{1}{2}T_v(S_0 - S_1)$$

$$I_{h,45°} = \frac{1}{2}T_h(S_0 - S_1) \qquad I_{v,45°} = \frac{1}{2}T_v(S_0 + S_1)$$

Taking the $I_h$ channels (i.e. taking measurements measured with the same polarization channel at different times), we can solve to acquire

$$S_0 = T_h(I_{h,0°} + I_{h,45°}) \qquad S_1 = T_h(I_{h,0°} - I_{h,45°})$$

Equivalently we could have used the $I_v$ channels to acquire

$$S_0 = T_v(I_{v,45°} + I_{v,0°}) \qquad S_1 = T_v(I_{v,45°} - I_{v,0°})$$

In both cases we can recover $S_0$ and $S_1$ which are scaled by a common quantity $T_{h/v}$. Since absolute count numbers are generally not of interest (in any case the counts reaching the detector will vary with measurement parameters), this allows us to recover the main quantities of interest i.e. the relative magnitudes of $S_0$ and $S_1$.

**Time Drift**

It is common to have some time-dependent intensity variation between temporally separated measurements. As both channels are measured simultaneously, this will only affect measurements taken at different waveplate orientations. This "drift" could therefore also incorporate changes in transmission caused by waveplate rotation itself (though for error cancelling, these must not be channel-dependent). For this, we would add drift factors $d_{0°}$ and $d_{45°}$ for the measurements taken at HWP angles 0° and 45°, respectively:

$$I_{h,0°} = \frac{1}{2}d_{0°}(S_0 + S_1) \qquad I_{v,0°} = \frac{1}{2}d_{0°}(S_0 - S_1)$$

$$I_{h,45°} = \frac{1}{2}d_{45°}(S_0 - S_1) \qquad I_{v,45°} = \frac{1}{2}d_{45°}(S_0 + S_1)$$

In this case, we can take the $I_{0°}$ channels (i.e. taking measurements measured at the same time with different polarization channels), to acquire

$$S_0 = d_{0°}(I_{h,0°} + I_{v,0°}) \qquad S_1 = d_{0°}(I_{h,0°} - I_{v,0°})$$

Equivalently we could have used the $I_{45°}$ channels to acquire

$$S_0 = d_{45°}(I_{h,45°} + I_{v,45°}) \qquad S_1 = d_{45°}(I_{v,45°} - I_{h,45°})$$

Where again we can recover $S_0$ and $S_1$ which are scaled by a common quantity $d_{0/45°}$ and therefore recover the main quantities of interest i.e. the relative magnitudes of $S_0$ and $S_1$.

**Combined Imperfections**

If both time drift and transmission imbalance is present, following the same treatment as before we end up with:

$$I_{h,0} = \frac{1}{2}d_{0°}T_h(S_0 + S_1) \qquad I_{v,0} = \frac{1}{2}d_{0°}T_v(S_0 - S_1)$$

$$I_{h,45} = \frac{1}{2}d_{45°}T_h(S_0 - S_1) \qquad I_{v,45} = \frac{1}{2}d_{45°}T_v(S_0 + S_1)$$

This time, it is clear that no combination of two measurements will cancel out errors.

Following Baguenard *et al.* we can calculate the average intensity difference between the two tracks across the two HWP positions as(*3*)

$$\Delta I = (I_{h,0°} + I_{v,45°}) - (I_{v,0°} + I_{h,45°})$$

$$= \frac{1}{2}[(d_{0°}T_h(S_0 + S_1) + d_{45°}T_v(S_0 + S_1)) - (d_{0°}T_v(S_0 - S_1) + d_{45°}T_h(S_0 - S_1))]$$

$$= \frac{1}{2}[(d_{0°}T_h + d_{45°}T_v + d_{0°}T_v + d_{45°}T_h)S_1 + (d_{0°}T_h + d_{45°}T_v - d_{0°}T_v - d_{45°}T_h)S_0]$$

And similarly the combined total intensity as

$$I_{total} = (I_{h,0°} + I_{v,45°}) + (I_{v,0°} + I_{h,45°})$$

$$= \frac{1}{2}[(d_{0°}T_h(S_0 + S_1) + d_{45°}T_v(S_0 + S_1)) + (d_{0°}T_v(S_0 - S_1) + d_{45°}T_h(S_0 - S_1))]$$

$$= \frac{1}{2}[(d_{0°}T_h + d_{45°}T_v - d_{0°}T_v - d_{45°}T_h)S_1 + (d_{0°}T_h + d_{45°}T_v + d_{0°}T_v + d_{45°}T_h)S_0]$$

Notably, $\Delta I$ and $I_{total}$ are not completely equal to $S_1$ and $S_0$ multiplied by some common factor (as was the case when time/transmission errors were considered in isolation), but retain components from both Stokes components. This admixture has the magnitude

$$d_{0°}T_h + d_{45°}T_v - d_{0°}T_v - d_{45°}T_h$$

and therefore should be small. However, it remains the case that when both time instability and channel transmission inhomogeneity are present, the $S_1$ component estimated through $\Delta I$ contains a term which is linearly proportional to $S_0$ (i.e. a polarization artifact). Less importantly (as usually we have $S_0 \gg S_1$), the $S_0$ component estimated through $I_{total}$ contains a term which is linearly proportional to $S_1$.

Following similar reasoning, an estimate for $S_1/S_0$ might be approached by calculating (similar to the g$_{lum}$ calculation of Baguenard(*3*))

$$\frac{\Delta I}{I_{total}} = \frac{(d_{0°}T_h + d_{45°}T_v + d_{0°}T_v + d_{45°}T_h)S_1 + (d_{0°}T_h + d_{45°}T_v - d_{0°}T_v - d_{45°}T_h)S_0}{(d_{0°}T_h + d_{45°}T_v - d_{0°}T_v - d_{45°}T_h)S_1 + (d_{0°}T_h + d_{45°}T_v + d_{0°}T_v + d_{45°}T_h)S_0}$$

Defining $\alpha = \frac{d_{0°}T_v + d_{45°}T_h}{d_{0°}T_h + d_{45°}T_v}$, this is simplified to

$$\frac{\Delta I}{I_{total}} = \frac{(1+\alpha)S_1 + (1-\alpha)S_0}{(1-\alpha)S_1 + (1+\alpha)S_0}$$

Which only simplifies to $S_1/S_0$ if $\alpha = 1$; this only holds if $d_{0°} = d_{45°}$ or $T_h = T_v$ i.e. a combination of both time drift and channel transmission imbalance is not present.

On the other hand, we could simply note that

$$\frac{I_{v,0°} \cdot I_{h,45°}}{I_{h,0°} \cdot I_{v,45°}} = \left(\frac{S_0 - S_1}{S_0 + S_1}\right)^2$$

which gives

$$\frac{S_1}{S_0} = \frac{1-C}{1+C}$$

Where $C = \sqrt{\frac{I_{v,0°} \cdot I_{h,45°}}{I_{h,0°} \cdot I_{v,45°}}}$ only contains directly measured quantities. Therefore, it is possible to extract out a fully error-corrected quantity $S_1/S_0$ even in the presence of combined time drift and channel transmission imbalance. Since $I_{total}$ should be a good approximation of $S_0$ in the limit $S_0 \gg S_1$ and $\alpha = \frac{d_{0°}T_v + d_{45°}T_h}{d_{0°}T_h + d_{45°}T_v} \approx 1$, we can the recover an approximation of $S_1$ as

$$S_1 \approx I_{total} \frac{1-C}{1+C}$$

Thereby recovering all the quantities of interest.

The difference between calculating $S_3/S_0$ through $\frac{\Delta I}{I_{total}}$ and $\frac{1-C}{1+C}$ is illustrated in the figure below, where "channel avg." refers to $\frac{\Delta I}{I_{total}}$ and "better error cancelling" to $\frac{1-C}{1+C}$. It can be seen that a very similar noise level and numerical value is obtained by either method.

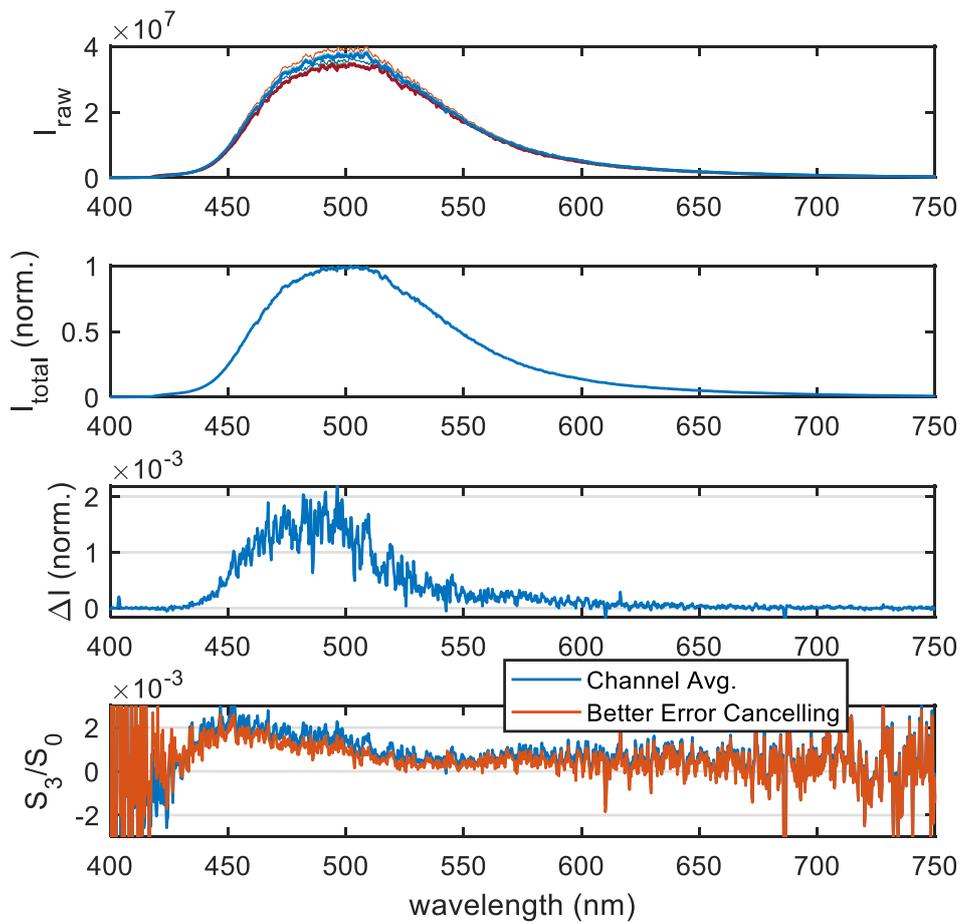

*Figure S5: $S_3$ measurement of r-BPC in toluene solution (excitation 343 nm, 200 fs, 12.5 kHz). The quantity $S_3/S_0$ in the bottom panel is calculated in two different ways with different degrees of error cancellation, as outlined in the text.*

## Section 3: Supplementary Data

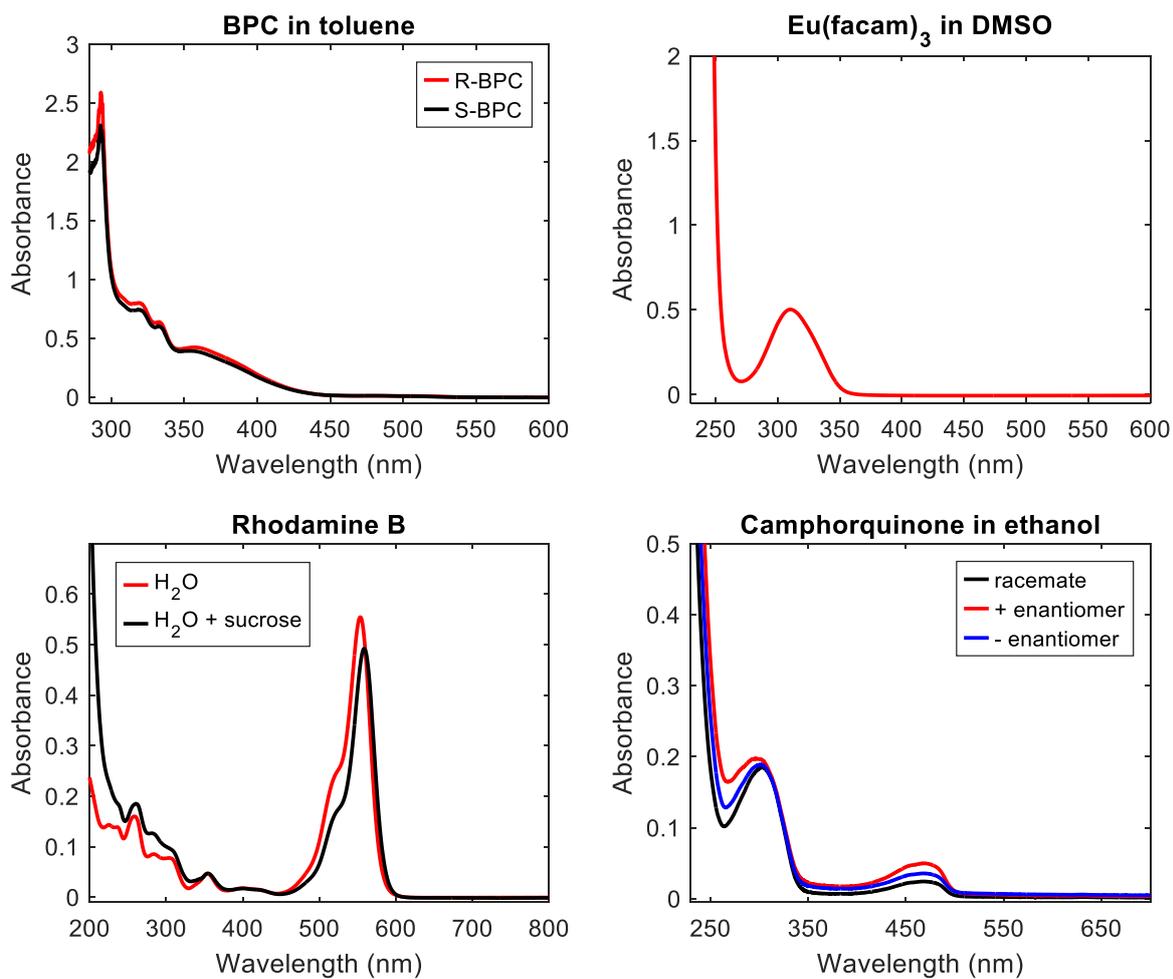

*Figure S6: UV-Visible absorbance spectra of the solutions used in this work.*

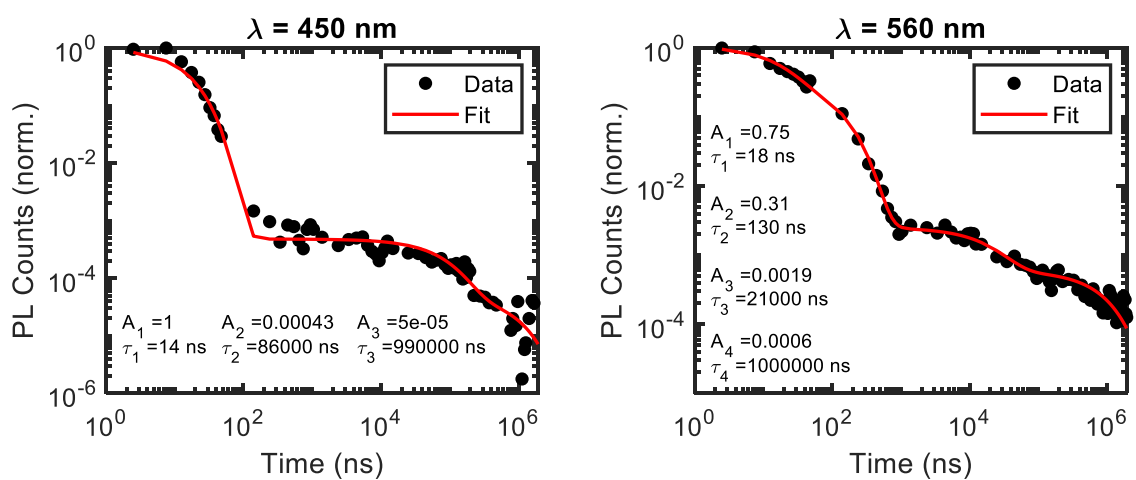

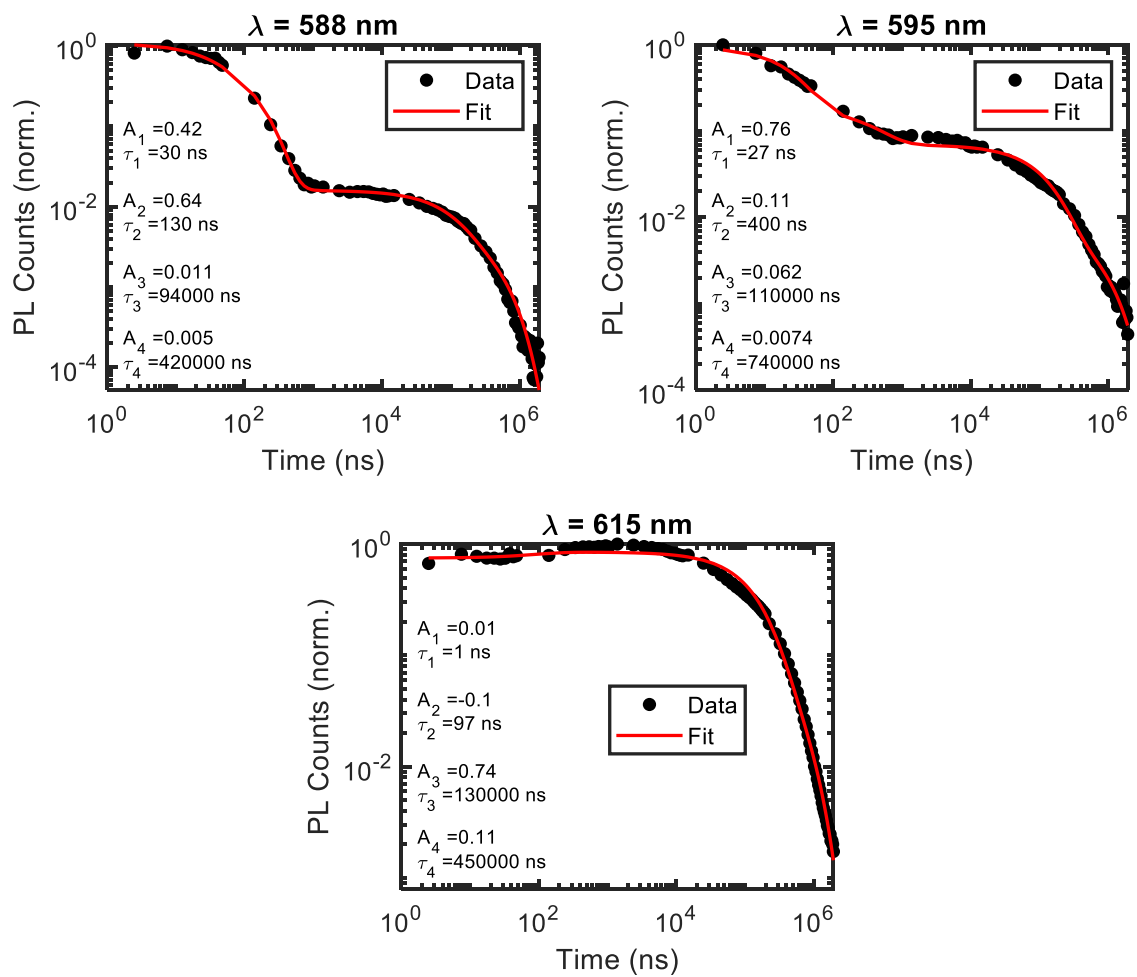

*Figure S7: Decay traces and fitted multi-exponential kinetics for non-polarization sensitive luminescence measurements of Eu[(+)-facam]$_3$ in DMSO at various wavelengths corresponding to spectral features of interest (excitation 343 nm, 200 fs, 500 Hz).*

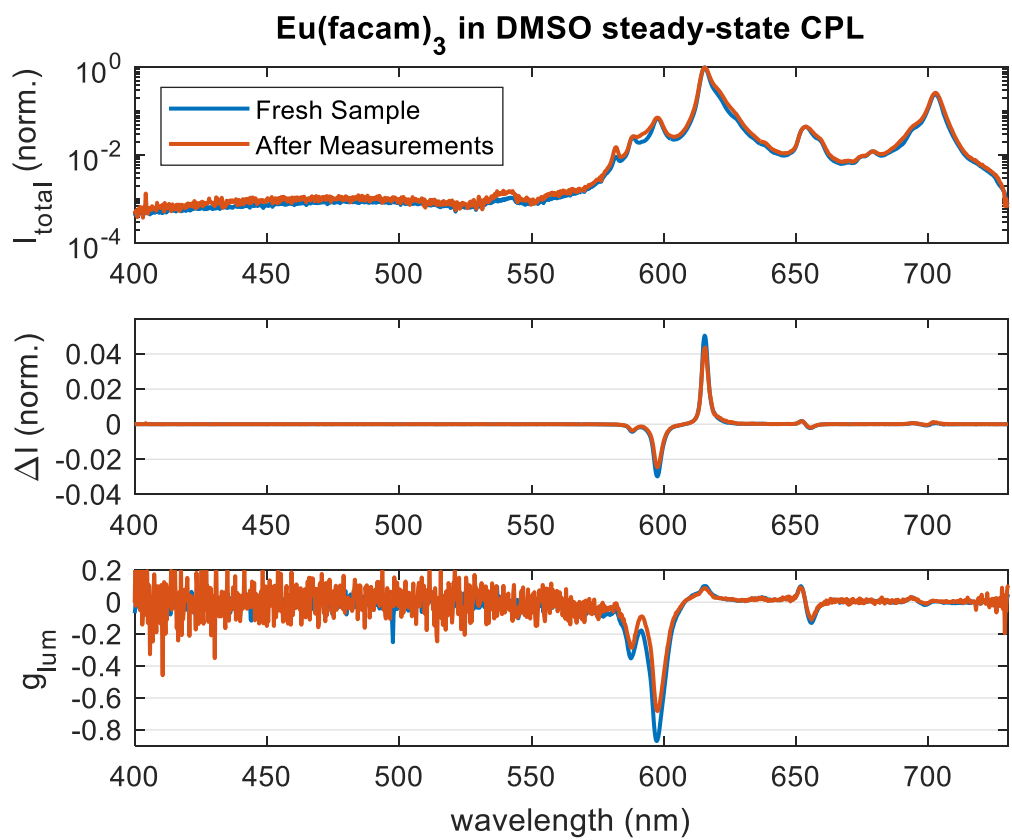

*Figure S8: Time-integrated (excitation 343 nm, 200 fs pulses, 50 kHz gate pulse width 20 µs covering full time between pulses) CPL measurements of Eu[(+)-facam]$_3$ in DMSO for a freshly prepared solution and after multiple hours of measurement time with pulsed 343 nm excitation, showing the decrease in CPL activity likely resulting from sample degradation.*

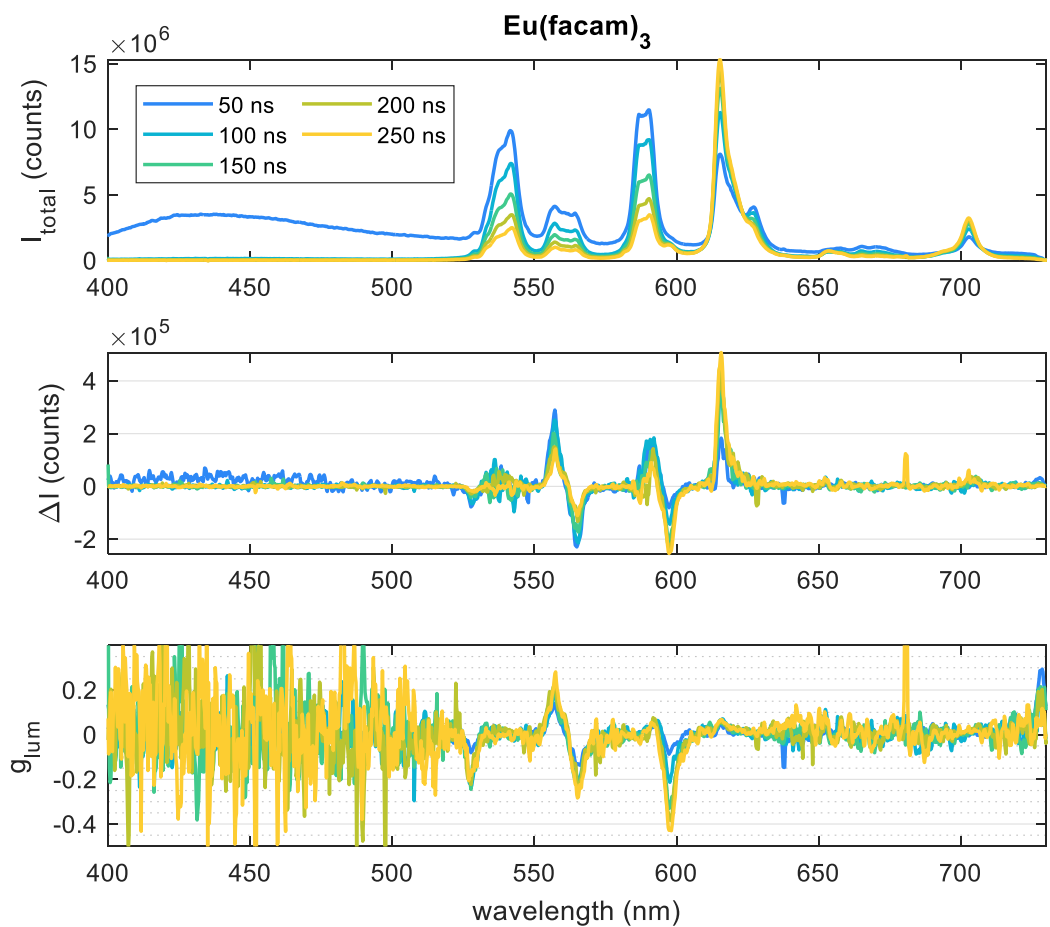

*Figure S9: Time-resolved CPL measurements of Eu[(+)-facam]₃ in DMSO with lower excitation frequency (excitation 343 nm, 200 fs, 500 Hz) to minimize wraparound emission.*

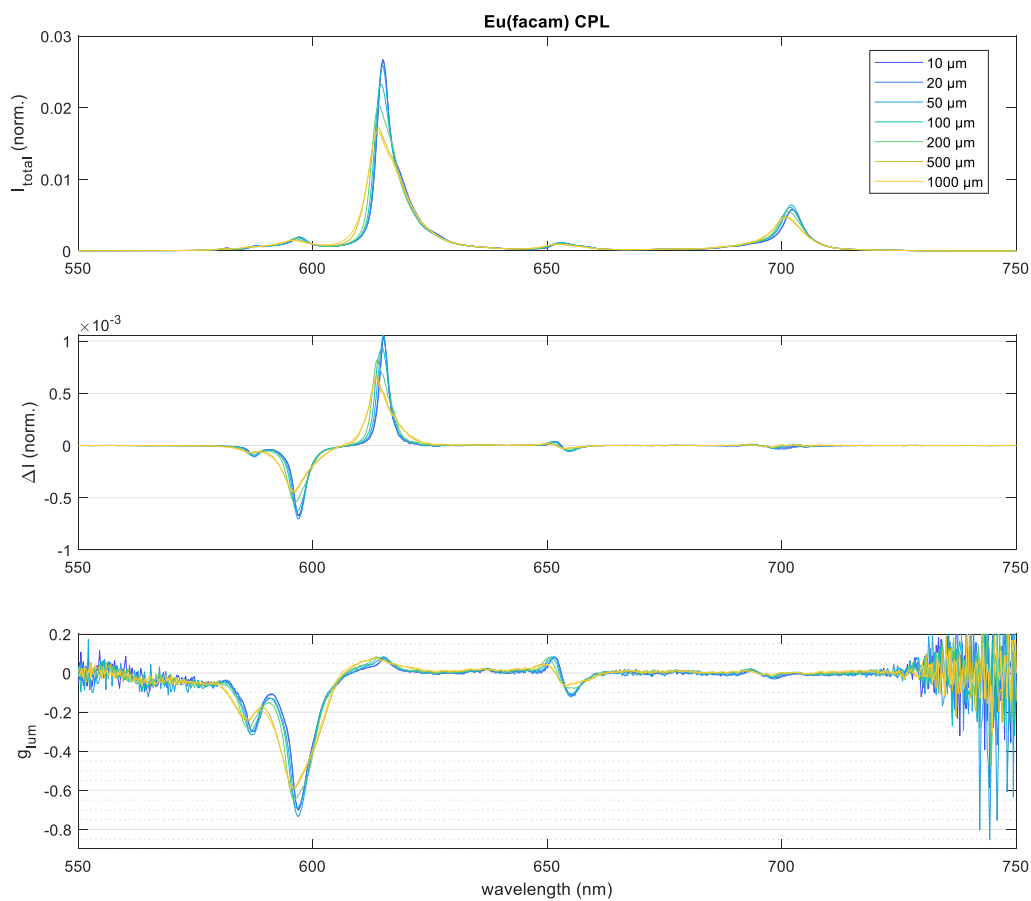

*Figure S10: Steady-state measurements of Eu[(+)-facam]$_3$ in DMSO with varying spectrograph slit widths. Note that as the slit size approaches the detector pixel size (13.5 μm) decreasing slit sizes becomes less effective at improving resolution.*

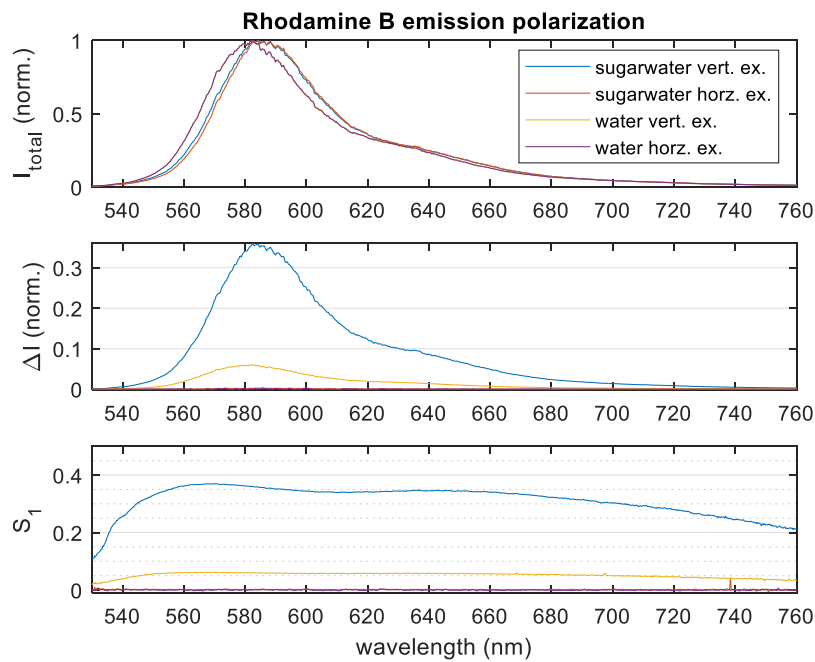

*Figure S11: $S_1$ Stokes measurement of Rhodamine B in aqueous solutions with horizontal and vertical excitation polarization (excitation 515 nm, 200 fs, 50 kHz).*

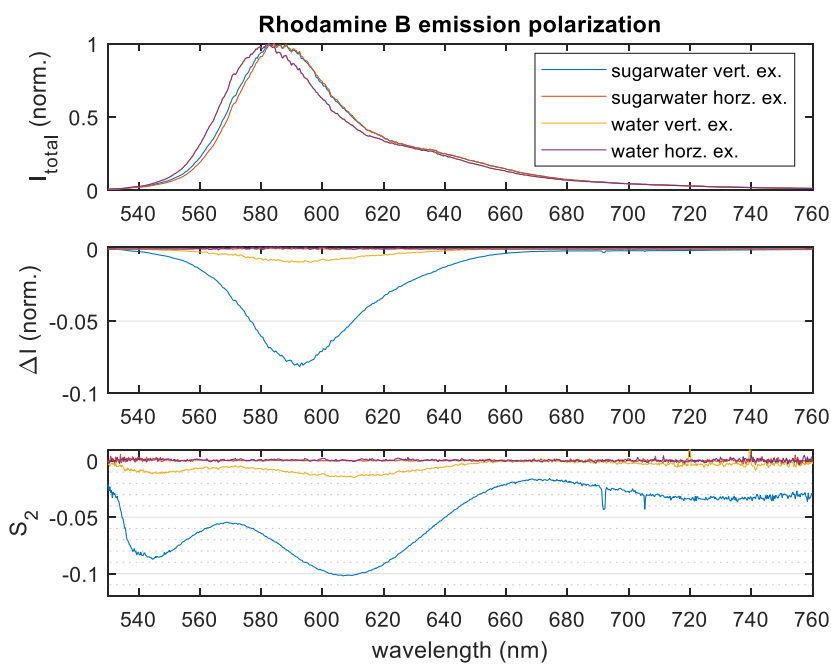

*Figure S12: $S_2$ Stokes measurement of Rhodamine B in aqueous solutions with horizontal and vertical excitation polarization (excitation 515 nm, 200 fs, 50 kHz).*

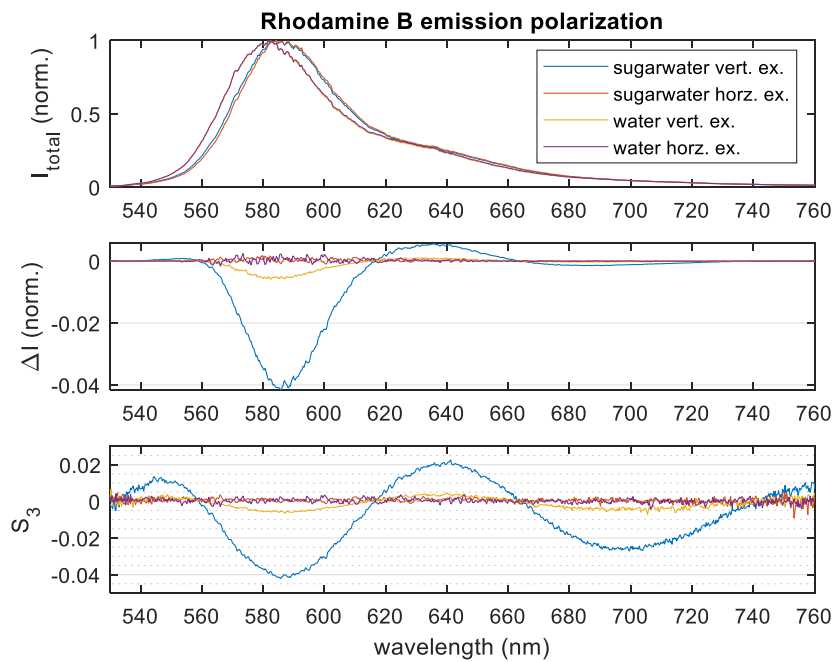

*Figure S13: $S_3$ Stokes measurement of Rhodamine B in aqueous solutions with horizontal and vertical excitation polarization (excitation 515 nm, 200 fs, 50 kHz).*

While sucrose is chiral, the polarization spectra retaining the same shape but different intensities with or without sucrose (Figures S10-S12) suggests that the origin of the apparent CPL signal is not any inherent chirality of the system but an artifact arising from the presence of linear polarization. This is further strengthened by the observation that emission linear polarization due to photoselection is much larger in magnitude when sucrose is present, likely due to the increased viscosity of the medium leading to slower orientational relaxation of dye molecules.

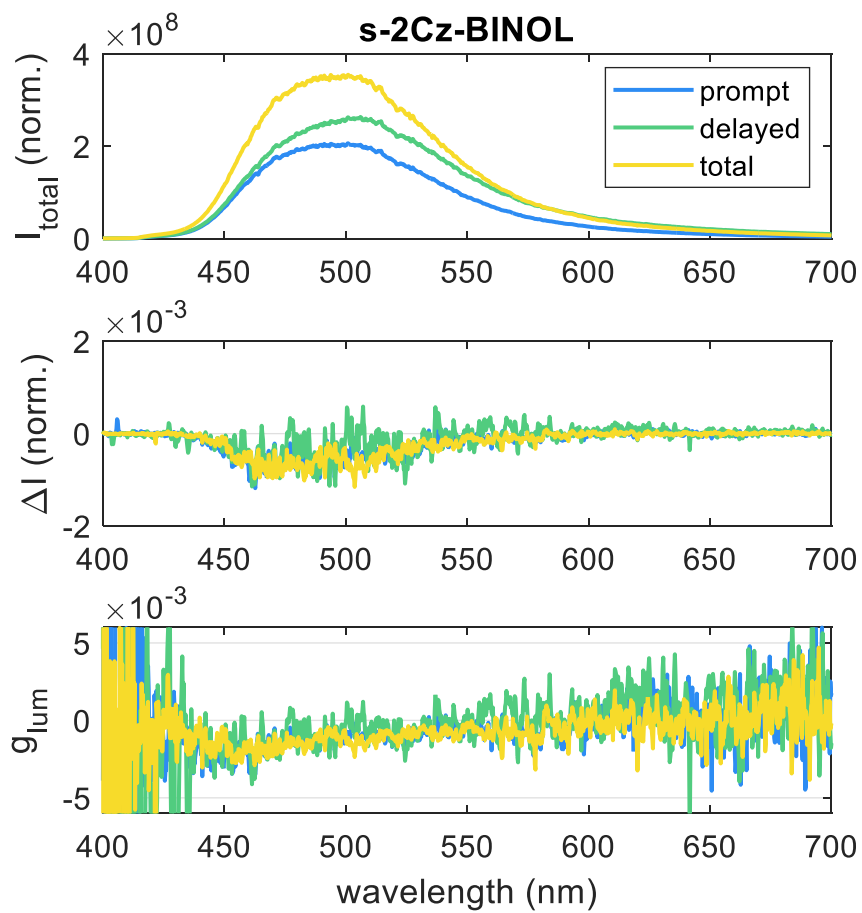

*Figure S14: Time-resolved CPL experiments of a chiral TADF-active dye S-BPH in toluene solution (excitation 343 nm, 200 fs, 12.5 kHz). Prompt' refers to the first 100 ns, 'Delayed' to approximately 500 ns–80 μs, and 'Total' to a gate covering the complete emission process (0–80 μs).*

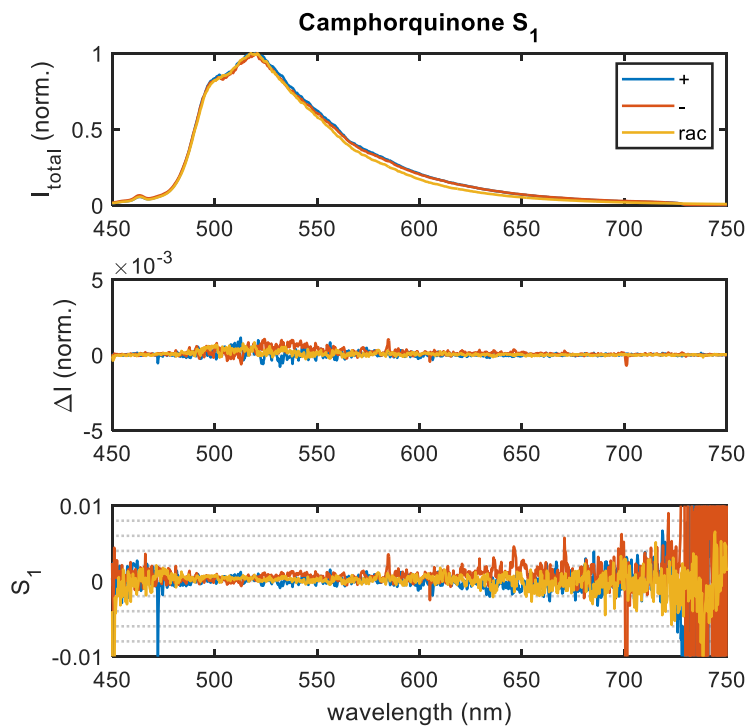

*Figure S15: $S_1$ Stokes measurement of camphorquinone (both enantiomers and racemate) in aqueous solutions with horizontal excitation polarization (excitation 405 nm, CW).*

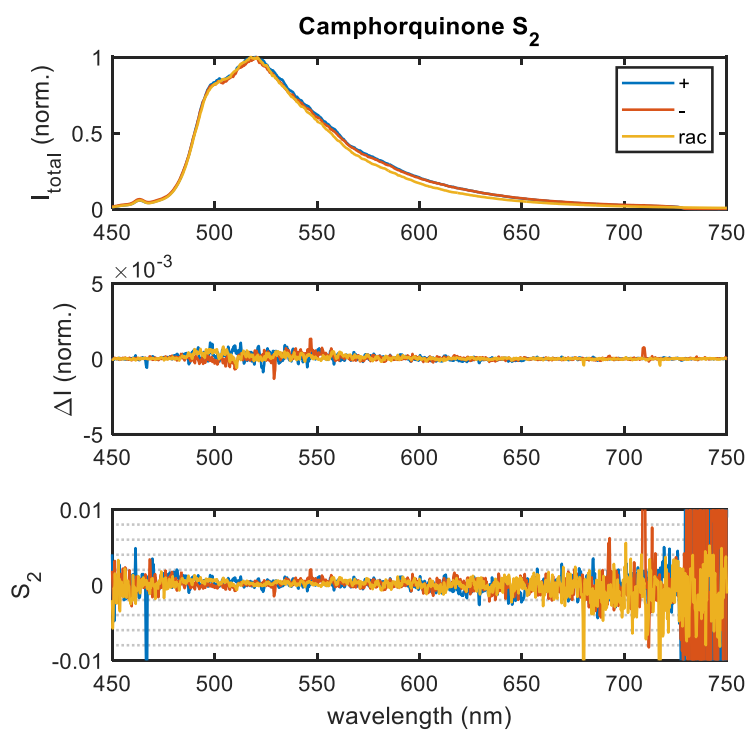

*Figure S16: $S_2$ Stokes measurement of camphorquinone (both enantiomers and racemate) in aqueous solutions with horizontal excitation polarization (excitation 405 nm, CW).*

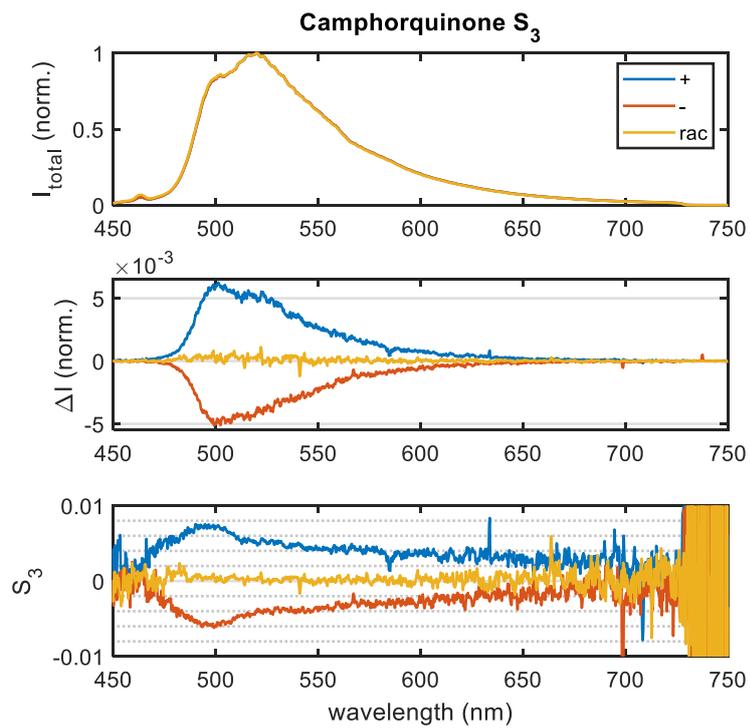

*Figure S17: S₃ Stokes measurement of camphorquinone (both enantiomers and racemate) in aqueous solutions with horizontal excitation polarization (excitation 405 nm, CW).*

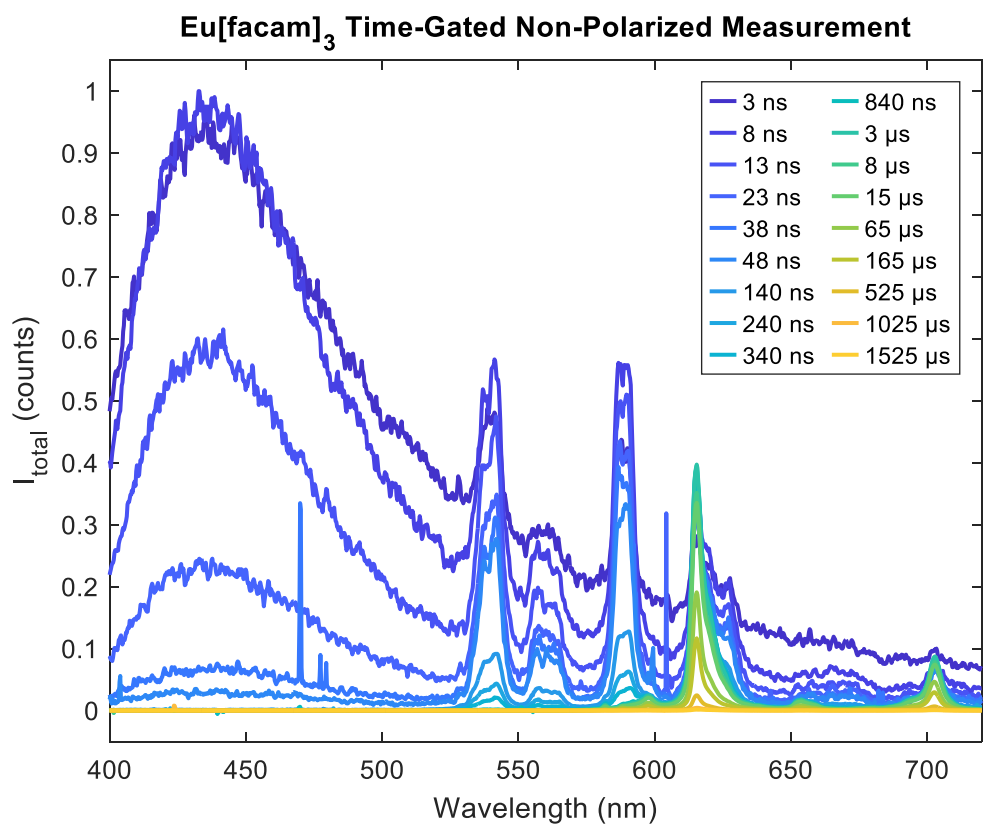

*Figure S18: Nanosecond time-resolved non-polarized luminescence of Eu[(+)-facam]3 in DMSO solution (excitation 343 nm, 200 fs, 500 Hz) with varying time bins. Data as presented in Figure 3 A of the main body, but without normalization to the peak value for each spectrum (values are normalized to the most intense pixel across all slices collected).*

## 4: Practical Considerations for the Setup Operation

### 4.1: Measurement Time and Setting Parameters

For accurate quantification of $g_{lum}$, a large number of photons at each wavelength must be counted. As a rough estimate, reaching a noise level of $10^{-4}$ requires $10^8$ photons in the ideal case where the only noise source is shot noise inherent to photon emission following Poisson statistics and signal-to-noise ratio scales with $\sqrt{N}$ for N photons counted. In practice, this will be an underestimate for our setup, as the intensified CCD detector has additional noise contributions compared to a photon-counting system (readout noise, shot noise in the dark signal, shot noise in the signal). We find that measuring such a large number of photons tends to be relatively time-consuming, owing to limits on detector readout rate, laser repetition rate and detector saturation effects. It is therefore important to properly define measurement settings for realistic measurement durations.

For faster measurements, we essentially want to maximize the number of charges collected and read out per second. This is most straightforwardly affected by excitation power (per pulse). Increasing the excitation laser (and thus gate pulse) repetition rate will increase signal strength, but normally we want to avoid wrap-around emission (i.e. emission that leaks into the photon collection after the next pulse) if possible. Beyond this, appropriate choice of time gates, exposure time and accumulations is important. Where appropriate, pixel binning is also a possible approach. For our purposes, we normally bin vertical pixels such that only two tracks are recorded and read out, which greatly increases detector readout rate compared to a full-sensor readout and reduces readout noise. In particular, increasing the intensifier gain will also amplify noise, and therefore should be the "last resort" parameter to increase count numbers.

### 4.2: Overexposure

While the gated iCCD detector does not suffer from photon pile-up in the same way as TCSPC measurements do, the detector will eventually saturate (pixels have a finite well depth), which together with the maximum readout speed and minimum exposure time provides a hard upper limit for data acquisition rate. Additionally, inherent to the iCCD device is a phosphor screen for converting incident photons into electrons for amplification and gating, and overexposing this phosphor will result in afterglow fading slowly over minutes to hours. Though this might affect both channels similarly, and normally would be a small effect compared to the real luminescence intensity, we nevertheless advise caution. Practically, we have found that phosphor overexposure is a concern before well depth when increasing incident light intensity.

### 4.3: Fixed Pattern Noise and Pixel Sensitivity

Another consequence of using a multichannel detector such as a CCD is that there is fixed pattern noise, that is, differences in count values read out by individual pixels when an identical amount of light is incident. This will be partially due to per-pixel variation of the dark signal, which will be temperature-dependent but otherwise fixed, and partially due to per-pixel variation in sensitivity due to manufacturing irregularities. The dark signal variation can be removed by background subtraction, and in principle the per-pixel sensitivity variation, like other channel-dependent sensitivity variations, should be corrected for (in the difference spectra and dissymmetry factor) by the two-step measurement procedure where a QWP rotation flips the horizontal/vertical channels. However, as discussed in the next section, this is not always the case.

### 4.4: Beam Drift and Interaction With Fixed Pattern Noise

We found that rotation of the QWP introduces a slight deflection of the beam. As such, the pixels which collect light in the second measurement are not exactly the same pixels as in the first measurement. Due to this, the track-swapping pixel sensitivity correction is incomplete, resulting in unexpectedly high static noise (i.e. consistent between measurements, assuming optics are not moved) approximately at the $10^{-2}$ level. This is sufficient to prevent high-sensitivity CPL measurements, regardless of the amount of charges accumulated. This is illustrated in Figure S19, where it can be seen that noise is consistently present in the same pixels even in separate measurements with different horizontal pixel binning values.

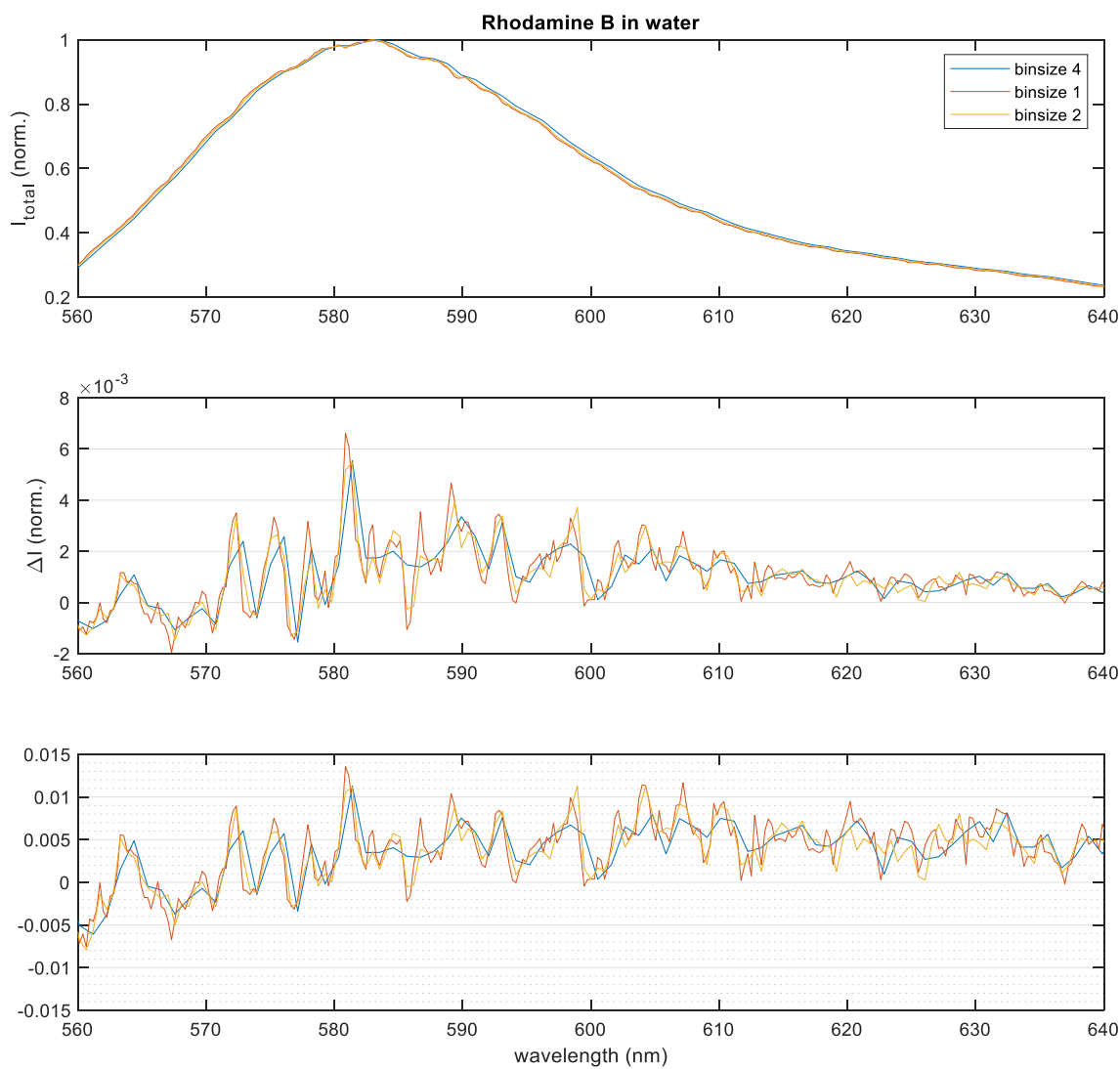

*Figure S19: CPL measurement of Rhodamine B in aqueous solution (excitation 515 nm, 200 fs, 50 kHz) with various horizontal pixel binning values. Note that this test data for the achiral dye has a remnant $g_{lum}$ offset.*

In the first instance, this can be corrected for by the somewhat crude but effective method of steering the excitation beam vertically after QWP rotation such that the same pixels are used to collect signal in both measurements. This is relatively simple, as the detector has a live imaging readout making alignment straightforward. However, it relies on the sample being very homogeneous (valid for solutions). Some amount of transmission change from beam steering is unavoidable, although this was found to work sufficiently well for high-accuracy measurements in our case.

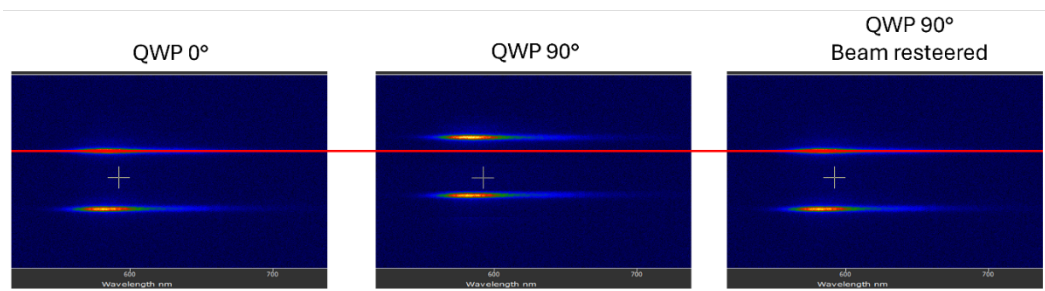

*Figure S20: Illustration of vertical beam drift on waveplate rotation. The spectral traces are translated vertically to different pixels when the beam drifts, resulting in incomplete error cancellation. By manually resteering the beam between measurements (done using the steering mirror immediately before the sample) this can be reversed. Images manipulated to exaggerate the effect for clarity.*

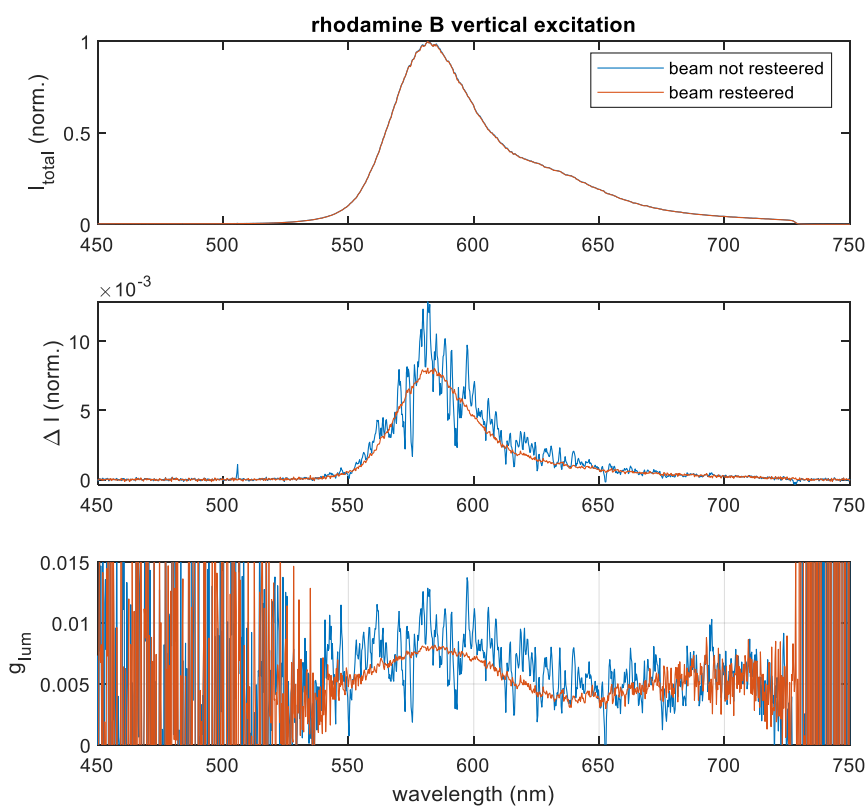

*Figure S21: CPL measurement of Rhodamine B in aqueous solution (excitation 515 nm, 200 fs, 50 kHz) measured with and without manual resteering of the beam between waveplate rotations, with otherwise exactly the same parameters. Note that this test data for the achiral dye has a remnant $g_{lum}$ offset.*

Manual beam resteering is illustrated in Figure S20, and the result on measurement "noise" in Figure S21. For exactly the same measurement parameters, the apparent noise level is massively improved by manually resteering the beam after waveplate rotation. This is because the "noise" without beam resteering is not true statistical noise, but rather static imperfections in the error cancellation caused by changing the pixels over which data is collected.

Horizontal beam deflection (orthogonal to the spectrograph entrance slit) is largely mitigated by using a 90⁰ excitation-collection geometry with a slow excitation focusing lens, such that the excitation beam forms a line through the sample. This has the added advantage of making alignment generally easier.

We have also found that the beam drift introduced by waveplate rotation can be mitigated by performing measurements with more than the minimum of two waveplate orientations. For CPL measurements, a total of four different orientations can be used, and for linear polarization measurements eight orientations are possible. Using multiple orientations helps due to the beam drift being at least partially reversed upon further rotation of the waveplate (as rotation by 360⁰ must return the beam to its initial position).

Besides mitigating beam deflection by tweaking optics, a possibly elegant and automatable approach to mitigate pixel sensitivity is the scanning multichannel approach. Here, instead of doing a single series of accumulations at a given grating position, a smaller number of accumulations is performed at multiple slightly offset grating positions (*51*) This approach has been successfully used to mitigate pixel noise in, for example, femtosecond stimulated Raman spectroscopy (*52*). As an added benefit, this will similarly smooth out any larger-scale sensitivity variations of the intensifier/detector, which normally have slightly lower sensitivity near the edge regions, without requiring a manual calibration file. However, transmission characteristics of optics will still contribute to spectral shape and require calibration to recover the original spectrum.

### 4.5: Slit Width

Slit width introduces the familiar tradeoff between intensity and resolution. For a narrow-line emission like that of $Eu^{3+}$, we recommend narrow (10-50 μm) slit widths to avoid smearing out features. This effect is shown in Figure S10, where the dissymmetry values for Eu(facam)$_3$ are shown to vary with slit width. For a pixel array detector, the pixel size usually introduces the lower limit to which slit size can improve resolution. Finer gratings allow for greater resolution at the cost of narrower bandpass (and, for holographic gratings, greater polarization sensitivity).

For broader spectral shapes, such as the organic molecules investigated here, a wider (100-200 μm) slit allows for more light incoupling and is advantageous.

### 4.6: Sample Degradation

As sample excitation happens by a pulsed laser (and often a pulsed UV laser), with accurate CPL requiring bright luminescence, sample degradation is a concern. Some degradation was observed in Eu(facam)$_3$, illustrated in $g_{lum}$ values being lower after prolonged laser exposure in Figure S8. To mitigate this, sample exposure should be minimized and fresh samples used. A flow cell might also be appropriate for sensitive applications.

## 5: Software Package

This section details our implemented software package and components critical to automate the measurement process in the proposed setup. Our proposed Algorithm 1 enables systematic acquisition of time- and polarization-resolved data, while the developed GUI simplifies user interaction with the system.

Automation brings numerous advantages, including ensuring reliability during measurements, saving time for performing measurements and improving reproducibility of results. One of our key motivations behind automation is to eliminate potential inaccuracies introduced by manual operations and to enhance user comfort. A measurement algorithm was developed for the given hardware configuration to time- and polarization-resolved measurements. The developed procedure is essential for automatically capturing data with our setup, specifically the glum and the full Stokes vector. For polarization-resolved measurements, the angle of the QWP and HWP are systematically varied, and measurements are accumulated. Time- and polarization-resolved measurements extend this approach to include time gating for capturing time-dependent information.

### 5.1 Algorithms

For the automated determination of the time-resolved stokes vector in our constructed setup we propose Time- and Polarization-Resolved Algorithm (TPRO). Pseudocode is provided in Algorithm 1, detailing the steps involved in capturing measurements under various conditions. Beyond pseudocode to make the procedure accessible from a high-level, we provide a detailed implementation on our GitHub repository (Link).

---
**Algorithm 1:** Time- and Polarization-Resolved Algorithm (TPRA)

**Require:** $\eta$: Number of accumulations, $q_{offset}$: QWP offset, $Q$: QWP angles, $h_{offset}$: HWP offset, $H$: HWP angles, $T$: desired time steps

1 adjust_spectrograph_settings();
2 $\mathcal{M}_{series} \leftarrow \{\}$
3 **foreach** $h_\psi$ in $H$ **do**
4 $\quad$ move_HWP_angle($h_\psi + h_{offset}$);
5 $\quad$ **foreach** $q_\theta$ in $Q$ **do**
6 $\quad\quad$ move_QWP_angle($q_\theta + q_{offset}$);
7 $\quad\quad$ $\mathcal{M}_{q\theta} \leftarrow \{\}$
8 $\quad\quad$ **foreach** $t$ in $T$ **do**
9 $\quad\quad\quad$ $\mathcal{M}_t \leftarrow \{\}$
10 $\quad\quad\quad$ **foreach** $n$ in $1$ to $\eta$ **do**
11 $\quad\quad\quad\quad$ $m_n \leftarrow$ perform_measurement();
12 $\quad\quad\quad\quad$ $\mathcal{M}_t \leftarrow \mathcal{M}_t \cup (m_n)$
13 $\quad\quad\quad$ $\mathcal{M}_{q\theta} \leftarrow \mathcal{M}_{q\theta} \cup (t, \mathcal{M}_t)$
14 $\quad\quad$ $\mathcal{M}_{series} \leftarrow \mathcal{M}_{series} \cup (h_\psi, q_\theta, \mathcal{M}_{q\theta})$

---

In total, six measurements are required which together yield the static stokes vector. First, the measurement of $S_1$ is carried out with the QWP fixed at 0°, while the HWP is set to 0° and 45°. Second, the measurement of $S_2$ is performed with the QWP at 0°, and the HWP rotated to 22.5° and 67.5°. Third, the $S_3$ (=$g_{lum} * S_0$) is performed with the HWP set at an arbitrary angle, along with the QWP orientations at 45° and -45°. Executing these measurements effectively switches the polarization tracks and minimizes the contributions from other polarizations, achieving the desired results. The procedure of stepping the QWP and HWP is described in lines 3-6. To extend the steady-state Stokes vector with time resolution, additionally, each of the six measurements must be performed at different time steps (line 8). These time steps are defined in terms of gate width and delay. The functionality and parameterization of gate width and delay have been explained in the main part. Finally, the loop in line 10 enables redundant measurements for each tuple (comprised of space and time) to minimize random errors based on accumulation.

Algorithm 1 can be extended by employing scanning multichannel detection (*51*) to improve the robustness against artifacts, namely, to reduce pixel sensitivity variations on the iCCD camera, as pointed out in the main part (Figure 1 c). This extension is depicted as pseudocode in Algorithm 2. To add scanning multichannel detection, each measurement n of the total number of accumulations N (line 10) is extended by g sub-measurements with slightly different grating positions (line 12).

**Algorithm 2: TPRA with Scanning Multichannel Detection**

**Require:** $\eta$: Number of accumulations, $q_{offset}$: QWP offset, $Q$: QWP angles, $h_{offset}$: HWP offset, $H$: HWP angles, $T$: desired time steps, $G$: grating positions

```
1  adjust_spectrograph_settings();
2  M_series ← {}
3  foreach h_ψ in H do
4      move_HWP_angle(h_ψ + h_offset);
5      foreach q_θ in Q do
6          move_QWP_angle(q_θ + q_offset);
7          M_qθ ← {}
8          foreach t in T do
9              M_t ← {}
10             foreach n in 1 to η do
11                 M_n ← {}
12                 foreach g in G do
13                     m_g ← perform_measurement();
14                     M_n ← M_n ∪ (g, m_g)
15                 M_t ← M_t ∪ (n, M_n)
16             M_qθ ← M_qθ ∪ (t, M_t)
17         M_series ← M_series ∪ (h_ψ, q_θ, M_qθ)
```

To perform the automated measurements on the real system, the proposed pseudocode in Algorithm 1 was implemented with Python. For hardware control, the code leverages the Python SDK from Andor Solis. While our pseudocode is generic and system-agnostic, the enumerated hardware components in Section 1 are required to successfully execute the provided Python implementation. The exact software packages that need to be installed to run the code are explained in detail on our GitHub repository. We would also like to point out that we have so far only implemented Algorithm 1 as Python code on our GitHub repository. We have already successfully implemented Algorithm 2 manually on our system, but we still need to write the Python code to automate it. Nevertheless, we consider it very useful for the community to share this extension in the form of pseudocode in Algorithm 2, so that they can implement/modify this extension themselves in advance if necessary or wait for our code extension.

**5.2 Graphical User Interface:**

Beyond our Python implementation of Algorithm 1 as a headless version, we have extended this with a Graphical User Interface (GUI) to facilitate user interaction with the system. The implementation serves as a frontend and was implemented with PyQT. The GUI allows users to conduct measurements but also to configure several measurement settings for the QWP, HWP, spectrograph and iCCD camera. A log section within the GUI provides real-time feedback on the measurement progress and information on the status of the system (e.g. current temperature). The GUI helps streamline the setup and measurement process and enhancing user convenience.